\documentclass[aps,prl,10pt,longbibliography,superscriptaddress,twocolumn,showpacs,oneside]{revtex4-2}

\usepackage[utf8]{inputenc}
\usepackage[american,british]{babel}
\usepackage[T1]{fontenc}
\usepackage{lmodern}
\usepackage{xcolor}

\usepackage{wrapfig}
\usepackage{graphicx}
\usepackage[caption=false]{subfig}

\usepackage{amsmath,amssymb,amsthm,mathrsfs,mathtools,amsfonts,physics,bm}
\usepackage{mathdots}
\usepackage{dsfont}
\usepackage[version=4]{mhchem}

\usepackage[colorlinks=true, linktocpage=true]{hyperref} 
\hypersetup{allcolors = {black}}

\usepackage{tikz}
\usetikzlibrary{shapes.geometric, arrows}

\usepackage{algorithm}
\usepackage{algpseudocode}
\usepackage{setspace}

\tikzstyle{Atensor} = [rectangle, rounded corners, minimum width=1cm, minimum height=1cm, text centered, draw=black]
\tikzstyle{Gtensor} = [circle, minimum width=1cm, minimum height=1cm, text centered, draw=black]
\tikzstyle{arrow} = [thick,->]

\usepackage{comment}
\usepackage[normalem]{ulem}  

\begin{document}

\title{Diverging conditional correlation lengths in the approach to high temperature}

\author{Jerome Lloyd}
\affiliation{Department of Theoretical Physics, University of Geneva, Geneva, Switzerland}

\author{Dmitry A. Abanin}
\affiliation{Department of Physics, Princeton University, Princeton NJ 08544, USA}

\author{Sarang Gopalakrishnan}
\affiliation{Department of Electrical Engineering, Princeton University, Princeton NJ 08544, USA}


\begin{abstract}
The Markov length was recently proposed as an information-theoretic diagnostic for quantum mixed-state phase transitions [Sang \& Hsieh, Phys.~Rev.~Lett.~134, 070403 (2025)]. Here, we show that the Markov length diverges even under classical stochastic dynamics, when a low-temperature ordered state is quenched into the high temperature phase. Conventional observables do not exhibit growing length scales upon quenching into the high-temperature phase; however, the Markov length grows exponentially in time. Consequently, the state of a system as it heats becomes increasingly non-Gibbsian, and the range of its putative ``parent Hamiltonian'' must diverge with the Markov length. From this information-theoretic point of view the late-time limit of thermalization is singular. We introduce a numerical technique for computing the Markov length based on matrix-product states, and explore its dynamics under general thermal quenches in the one-dimensional classical Ising model. For all cases, we provide simple information-theoretic arguments that explain our results. 
\end{abstract}
\maketitle

\emph{Introduction.---} 
The error correction threshold for topological quantum codes, such as the toric code~\cite{dennis2002topological}, exemplifies a class of phenomena named mixed-state phase transitions~\cite{PRXQuantum.5.020343, PhysRevB.110.085158, PhysRevX.15.011069, coser2019classification}. Mixed-state phase transitions differ from conventional thermal or quantum phase transitions in fundamental ways. If one applies noise at some finite rate $\gamma$ to a system initialized in one of the logical states of the toric code, the logical information ceases to be retrievable at some \emph{finite} time. Since conventional correlation functions obey light-cone bounds even in open systems~\cite{PhysRevLett.104.190401}, they cannot diverge at mixed-state transitions. Instead, mixed-state phases and phase transitions must be diagnosed by unconventional information-theoretic order parameters~\cite{PRXQuantum.5.020343}: the significance of these quantities is that they bound the performance of tasks like error correction. Identifying such order parameters, and developing a classification of mixed-state phases~\cite{coser2019classification, PhysRevX.14.041031}, has been a central theme in recent work. 
Very recently it was proposed~\cite{shengqi} that the appropriate information-theoretic length-scale that diverges at mixed-state transitions is the \emph{Markov length}, associated with the decay of conditional mutual information (CMI). The CMI is not subject to light-cone bounds~\cite{PRXQuantum.4.040332, PhysRevA.110.032426, PhysRevLett.133.200402}, so its range can diverge even at finite times~\cite{PhysRevLett.132.030401, PhysRevX.12.021021, PRXQuantum.3.040337}. A diverging Markov length signals that errors cannot be corrected by any quasilocal strategy~\cite{shengqi, sang2025mixed}. 

\begin{figure}[!b]
\begin{center}
\includegraphics[width = 0.48\textwidth]{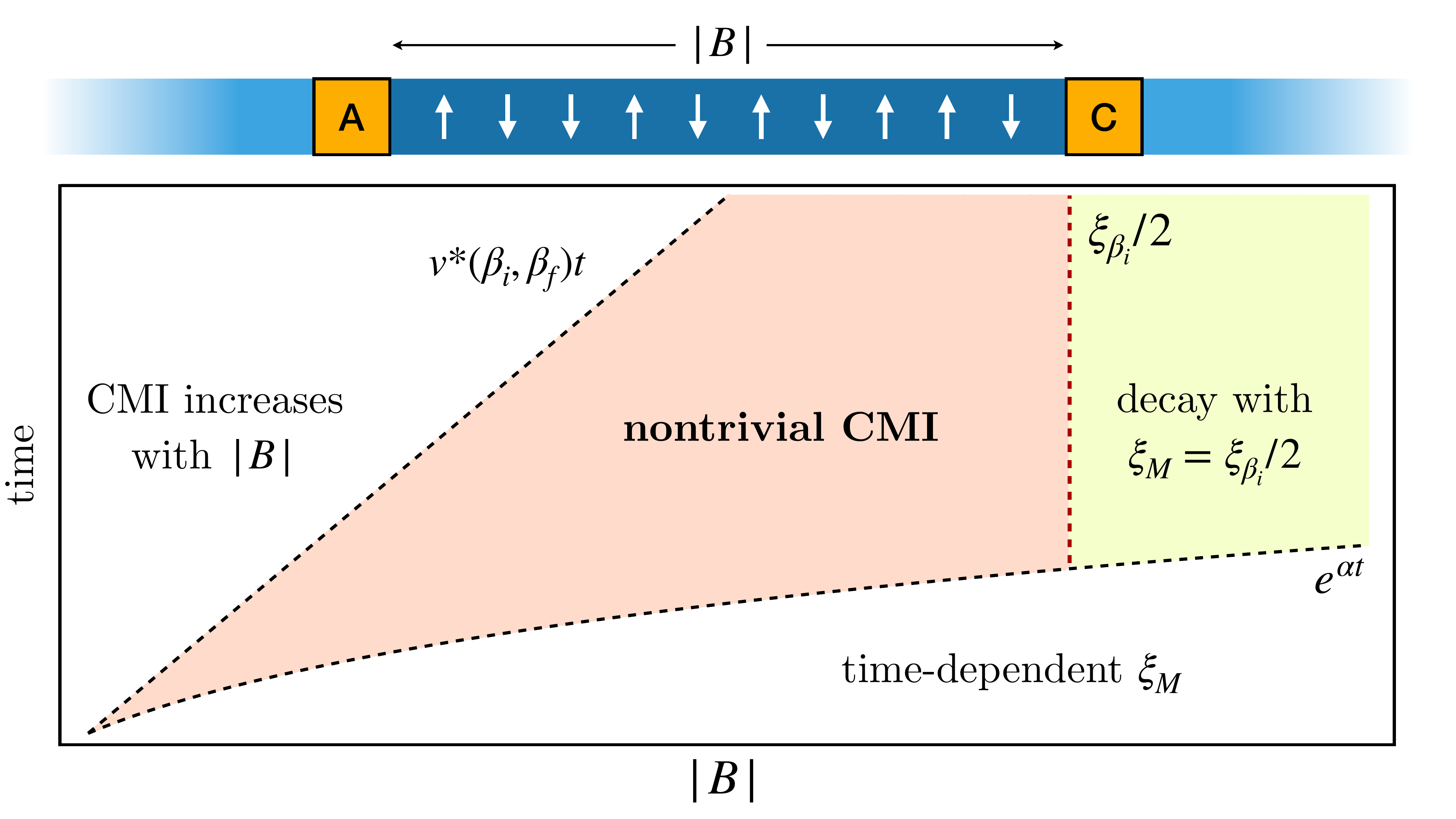}
\caption{Upper panel: conditional mutual information $I(A:C|B)$, where $B$ separates regions $A$ and $C$: i.e., there are no direct interactions between $A$ and $C$. The spins in $B$ are fixed to some configuration $\vec{\sigma}_b$ (drawn from the marginal $\pi(B)$) and $I(A:C)$ is computed in the conditional distribution $\pi(A:C|\vec{\sigma}_b)$. Lower panel: schematic diagram of the spacetime regimes of CMI. The CMI vanishes at short distances (where the dominant correlations are thermal) $|B| \ll t (\xi_{\beta_i} - \xi_{\beta_f})/(\xi_{\beta_i} \xi_{\beta_f})$, where $\xi_{\beta_i}, \xi_{\beta_f}$ are thermal equilibrium correlation lengths at the initial and final temperatures, and also at long distances $|B| \gg \min(e^{\alpha t}, \xi_{\beta_i}/2)$; it peaks in the intermediate region, shaded in orange.}
\label{schematic}
\end{center}
\end{figure}

So far, the CMI and Markov length have been considered primarily as static properties of a phase. 
In the present work, we address the \emph{dynamics} of the Markov length, focusing on the simple case of classical stochastic dynamics. In particular, we consider how the Markov length evolves under classical heating: i.e., we consider a system that is initially in equilibrium at inverse temperature $\beta_i = 1/T_i$ and evolve it under stochastic Glauber dynamics that obeys detailed balance and thus drives the system to equilibrium at a higher temperature, $\beta_f< \beta_i$. 
In the classical context, the CMI is simple to define: Consider a one-dimensional system, tripartitioned into regions $A, B, C$ (Fig.~\ref{schematic}). The CMI $I(A:C|B)$ is defined as the mutual information between $A$ and $C$, conditioned on the state of $B$ (Eq.~(\ref{eq:CMIsum})). A key property of the classical CMI is the Hammersley-Clifford theorem~\cite{hammersley1971markov}: for the geometry shown in Fig.~\ref{schematic}, where $B$ separates $A$ and $C$, $I(A:C|B) = 0$ iff the probability distribution $\pi(ABC) \propto \exp(- H_{AB} - H_{BC})$, where $H_{AB}$ ($H_{BC}$) are functions of the marginals $\pi(AB)$ ($\pi(BC)$). Vanishing CMI is thus a hallmark of Gibbs states of local Hamiltonians: if a distribution has long-range CMI it cannot be a thermal state for any local Hamiltonian. If $I(A:C|B)$ decays with the radius of $B$ as $\exp(-|B|/\xi_M)$, the coefficient $\xi_M$ is called the Markov length. 

Our central result is that the approach to high temperatures is singular for any $\beta_f < \beta_i$, if one considers the Markov length. This singular behavior is most striking when $\beta_i$ corresponds to an ordered low-temperature state and $\beta_f$ is in a disordered high-temperature state. In this case, the Markov length grows without bound as $\xi_M(t) \sim \exp(t)$, while the long-distance behavior of the CMI is $I(A:C|B) \sim \exp(-t) \exp(-|B|/\xi_M(t))$: thus, corrections to the thermal equilibrium state at late times are exponentially small but also exponentially non-local. At late finite times, $\xi_M(t)$ is large; precisely at infinite time, $\xi_M = 0$ by the Hammersley-Clifford theorem, since the steady state is in thermal equilibrium. Remarkably, the late-time limit is singular even when both $\beta_i$ and $\beta_f$ are in the same phase: here, once again, the overall magnitude of long-range CMI decays exponentially in time, but at late times its range approaches $\xi_M\sim\xi_{\beta_i}/2$, where $\xi_{\beta_i}$ is the thermal correlation length at the lower temperature~\cite{zhang2025conditional}. Again, this late-time limit of $\xi_M$ disagrees with the exact steady-state value $\xi_M = 0$. We argue for these results on general physical grounds, and substantiate our arguments with tensor-network based simulations of the heating dynamics in a one-dimensional Ising model. In this framework, we show that $\xi_M$ can be interpreted as the Lyapunov gap of a spatial transfer matrix---a result that might be of independent interest.

\emph{Depolarizing ground states}.---We start with the simplest case, namely a quench from zero to infinite temperature. Consider 
the ferromagnetic Ising model, defined by the Hamiltonian $
    H(\vec{\sigma}) = -\sum_{i=1}^{L-1} \sigma_i \sigma_{i+1}$, 
where $\sigma_i = \pm 1$ are Ising spin variables and $\vec{\sigma} = (\sigma_1,\ldots,\sigma_L)$ the spin configuration. The ground state is given by the symmetric combination (`classical cat' state)
    $\theta = \frac{\ket{\Uparrow}+\ket{\Downarrow}}{2}$,
where e.g. $\ket{\Uparrow }$ is the state with all spins aligned in the `up' direction. (In what follows we will use the ket notation for classical probability distributions.) Physically, the classical cat state is one in which the observer does not know the sign of the ground state magnetization $m_0$. The ground state manifestly satisfies $I(A:C|B) = 0$ for \emph{any} tripartition. 

\begin{figure*}[!t]
    \centering
     \includegraphics[width=1\linewidth]{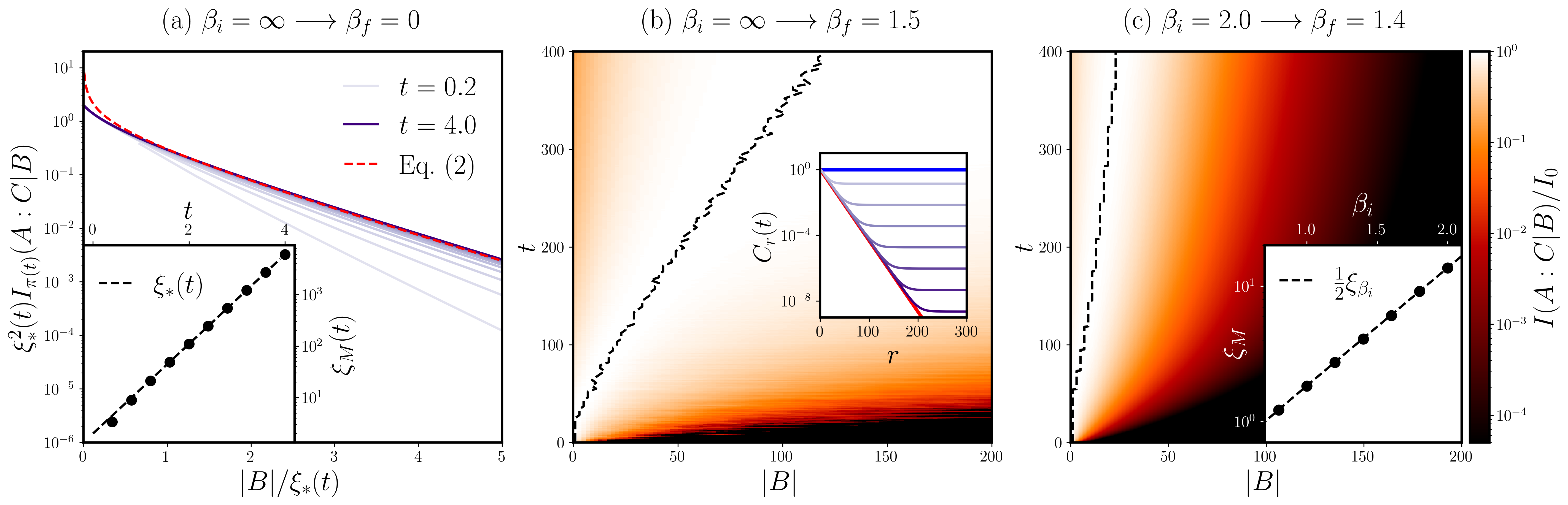}
     \caption{(a) \emph{Ising ground state under depolarization ---}  we plot the decay of CMI vs.~size of the $B$ region, $|B|$, showing a scaling collapse w.r.t.~the late-time Markov length prediction $\xi_*(t) = 2e^{2t}$. Colour transitions from light to dark purple denote increasing times; the dashed red line is the approximate solution Eq.~(\ref{eq:collapse}). In the inset we compare numerically extracted time-dependent Markov length (using a fit $\frac{1}{2}I(A:C|B)\big[\xi_*^2(t)\sqrt{6|B|/\xi_*(t)}\big] = e^{-|B|/\xi_M(t)}$) with the late-time prediction $\xi_M(t) = \xi_*(t)$. (b) \emph{Ising ground state under Glauber noise} --- heatmap (logarithmic colour scale) for space-time dependence of CMI under heating to final temperature $\beta_f = 1.5$. We normalise the CMI by the maximum $I_0 = \text{max}_{|B|}I(A:C|B)$.  The black dashed line denotes the position of the maximum. At late times, the decay of the CMI past the maximum is imperceptible, consistent with a diverging Markov length. \emph{Inset:} evolution of spin-spin correlation function $C_r(t) = \langle \sigma_0\sigma_r\rangle_{\pi(t)}$. The early-time/long-distance behaviour is set by the initial thermal correlation length (blue line), while the late-time/short-distance behaviour is set by the final correlation length (red line). (c) \emph{Thermal state under Glauber noise} --- space-time dependence of CMI, as in (b), for $\beta_i=2$ and $\beta_f=1.4$. At late times, the CMI decays with a Markov length set by the initial state correlation length. \emph{Inset:} Late-time numerical value of Markov length vs.~initial state temperature $\beta_i$. Dashed line shows $\xi_M = \xi_{\beta_i}/2$, where $\xi_{\beta} = -1/\log\tanh\beta$.}\label{fig:glauber}
\end{figure*}

We now act on the ground state with depolarizing noise for a time $t$. We define the spin-flip probability $p(t) = (1-m(t))/2$, where $m(t) = e^{- t}$ is the decaying magnetisation starting from the all-up state, and the noise channel is defined by $\mathcal{E}_{p(t)}(\pi)= \prod_i \mathcal{E}_{p(t)}^{(i)}(\pi)$, with $\mathcal{E}^{(i)}_p$ the binary symmetric channel acting on the $i$-th spin:
\begin{equation}\label{eq:depolchannel}
   \mathcal{E}_p^{(i)}(\sigma_i|\sigma_i') = (1-p) \delta_{\sigma_i,\sigma'_i} +p(1-\delta_{\sigma_i,\sigma'_i}).
\end{equation}
We denote the time-evolved state as $\ket{\pi(t)}$; quantities evaluated in this state will have a subscript $\pi(t)$. Since the noise acting on different spins is uncorrelated, any product distribution remains product. The classical cat state is not a product distribution, but has a simple decomposition in terms of product distributions, so for this initial state $\ket{\pi(t)}$ takes a simple form, $\ket{\pi(t)} = \frac{1}{2} (\otimes_i \ket{\pi^+_i} + \otimes_i \ket{\pi^-_i})$, where $\ket{\pi^\pm_i} = (1-p^\pm(t))\ket{\uparrow_i}+p^\pm(t)\ket{\downarrow_i}$ is a single-site distribution with $p^\pm(t) = \frac{1}{2} (1 \pm m(t))$. Given this explicit representation of $\ket{\pi(t)}$, we can straightforwardly evaluate $I_{\pi(t)}(A:C|B)$. For specificity we choose a tripartition in which $A$ and $C$ are single spins, separated by a contiguous region comprising $|B|$ spins. 

The CMI can be evaluated exactly numerically (see Supplemental Material \cite{SM}), as well as at late times in the scaling limit $m^2(t)|B|\gg1$, where a saddle point calculation gives
\begin{equation}\label{eq:collapse}
    I(A:C|B) \approx \frac{2}{\xi_*^2(t)\sqrt{6|B|/\xi_*(t)}}e^{-|B|/\xi_*(t)}.
\end{equation}
Here, $\xi_*(t) = 2 /m^{2}(t) = 2e^{2t}$ is the late-time Markov length, which therefore grows exponentially with time. In Fig.~\ref{fig:glauber}(a) we plot the scaling collapse $\xi_*^2(t)I(A:C|B) = f(|B|/\xi_*(t))$, for 20 equally space time points in the range $ t\in [0.2,4]$, and compare to the prediction Eq.~(\ref{eq:collapse}). Note that Eq.~(\ref{eq:collapse}) remains consistent with the Hammersley-Clifford theorem. While the Markov length diverges at infinite time, when the system has reached the infinite temperature equilibrium state, the prefactor $\xi_*^{-2}$ ensures the total magnitude of the CMI is simultaneously vanishing at this point.  

The behavior of Eq.~\eqref{eq:collapse} has a straightforward information-theoretic interpretation. It is convenient to use the identity $I(A:C|B) = I(A:BC) - I(A:B)$, i.e., the CMI is the additional information gained about $A$ if one expands $|B|$ by one site. The only information about $A$ that can be inferred from the rest of the system is the sign of the global magnetization $m$. For any nonzero $m$, when $|B| \gg 1/m^2$, the sign of $m$ can be reliably inferred by majority vote on $B$. Thus, $I(A:B) \approx m^2$, and is insensitive to the size of $B$ (up to exponentially small corrections); accordingly, $I(A:C|B)$ is also exponentially small. On the other hand, when $|B| \ll 1/m^2$, by the central limit theorem the probability of guessing $\mathrm{sign}(m)$ correctly from $B$ scales as $m \sqrt{|B|}$. Therefore, the correlation between the net magnetization on $B$ and on $A$ scales as $m^2 \sqrt{B}$, from which it follows that $I(A:B) \sim m^4 |B|$ and $I(A:C|B) \sim m^4$. Although we considered a specific one-dimensional geometry, the logic here is completely general and applies to any initial state that spontaneously breaks a discrete symmetry, in any spatial dimension: information-theoretically, what these phases have in common is that they are repetition codes encoding $O(1)$ logical bits, and (in the limit of a large system) the repetition code remains decodable for \emph{any} finite $t$. Our results show that while the Markov length grows exponentially quickly, there is no true divergence at finite times, since the magnetisation remains finite. Rather, the mixed state phase transition for this repetition code occurs precisely at $t=\infty$.

\emph{Glauber dynamics via MPS.---} We now turn to the general case where the initial and/or final temperatures are finite. In the remainder of the paper we restrict attention to the 1d classical Ising model. In this setting, a standard choice for stochastic dynamics that obeys detailed balance and converges to a steady-state Gibbs distribution at inverse temperature $\beta$ --- i.e., $\theta_{\beta} \equiv \exp(-\beta H) / \mathcal{Z}$, where $\mathcal{Z}$ is the partition function --- is Glauber dynamics \cite{glauber1963time}. For numerical simplicity we consider a discrete-time noise channel
\begin{equation}
\label{eq:thermalupdate}
    \ket{\pi(t+1)} = \Phi_{\beta}\ket{\pi(t)} = \Phi^{\text{odd}}_{\beta}\circ\Phi^{\text{even}}_{\beta}\ket{\pi(t)},
\end{equation}
where first even-numbered spins are allowed to flip, and then odd-numbered spins. The channel $\Phi^{\text{even}}_{\beta}$ is given by
\begin{equation}
    \Phi^{\text{even}}_{\beta}(\vec{\sigma}|\vec{\sigma}') = \sum_{\vec{\tau}} \prod_{i\ \text{even}}  \Delta_{\tau_{i-1},\tau_{i}}^{\sigma_{i-1},\sigma'_{i-1}}W^{\sigma_i,\sigma'_i}_{\tau_i,\tau_{i+1}}(\beta'),
\end{equation}
with tensors 
\begin{equation}\label{eq:glauberweights}
    W^{\sigma,\sigma'}_{\tau,\tau'}(\beta) = (1-\alpha)\delta_{\sigma,\sigma'}+\alpha \frac{\exp(\beta \sigma(\tau+\tau'))}{\sum_{\sigma''} \exp(\beta \sigma''(\tau+\tau'))},
\end{equation}
\begin{equation}\label{eq:glauberweights2}
  \Delta_{\tau,\tau'}^{\sigma,\sigma'}  = \delta_{\sigma,\sigma'}\delta_{\tau,\tau'}\delta_{\sigma,\tau'},
\end{equation}
and $\delta_{\sigma,\sigma'}$ the Kronecker delta. Parameter $\alpha$ acts as the timestep, with the continuous-time channel approached as $\alpha \to 0$. The $\Phi^{\text{odd}}_{\beta}$ channel is defined analogously. Our choice of weights in Eq.~(\ref{eq:glauberweights}) corresponds to the discrete-time version of the thermalization process introduced in Glauber's paper \cite{glauber1963time}: the weights satisfy the detailed balance property,
\begin{equation}
    e^{-\beta\sigma'(\tau+\tau')}W^{\sigma,\sigma'}_{\tau,\tau'}(\beta) = e^{-\beta\sigma(\tau+\tau')}W^{\sigma',\sigma}_{\tau,\tau'}(\beta),
\end{equation}
which ensures that the fixed point of the $\Phi_\beta$ channel is given by the thermal state $\theta_\beta$. 

In general, Glauber dynamics is not exactly solvable (although the correlation functions can be calculated explicitly for the 1d Ising model \cite{glauber1963time, SM}), and we must resort to approximate numerical methods. Traditionally, Glauber dynamics has been studied using the toolbox of classical Monte Carlo methods. In order to compute the CMI, however, we need access to an explicit representation of the probability distribution $\pi(t)$, rather than samples as in Monte Carlo methods. We therefore employ matrix product state (MPS) simulations to evolve an approximation to the full probability distribution at every time step, encoding the correlations of the state in local tensors $X^{[j]}$. This method has been heavily developed in the context of quantum systems, and recently applied to classical dynamical problems \cite{johnson2010dynamical, johnson2015capturing, banuls2019using, causer2022finite, gu2022tensor}. For details on the MPS implementation we refer to \cite{SM}. Importantly, the initial thermal state can be encoded as an MPS state (with bond dimension 2), and the update step Eq.~(\ref{eq:thermalupdate}) efficiently implemented. We use the framework of uniform MPS throughout, which allows us to work directly in the thermodynamic limit and store a single translationally-invariant tensor $X^{[j]} = X$. 

Our main technical result is an efficient method to estimate the CMI within the MPS framework. A brute force method, first computing the subsystem entropies and then employing the formula $I(A:C|B) = S_{AB}+S_{BC}-S_B-S_{ABC}$, is impractical to study the long distance decay of the CMI, since the complexity of evaluating the entropy grows as $2^{|B|}$. Instead, we use the CMI formula
\begin{equation}\label{eq:CMIsum}
    I_{\pi(t)}(A:C|B) = \sum\nolimits_{\vec{\sigma}_b} \pi_B(\vec{\sigma}_b) I_{\pi(t)}(A:C|B=\vec{\sigma}_b),
\end{equation}
where $\vec{\sigma}_b = (\sigma_{b,1},\ldots,\sigma_{b,|B|})$ denotes a spin configuration on the region $B$, $\pi_B(\vec{\sigma}_b)$ is the corresponding marginal probability, and $I_{\pi(t)}(A:C|B=\vec{\sigma}_b)$ is the MI between $A$ and $C$ in the state conditioned on the outcome $\vec{\sigma}_b$. We then make use of the fact that configurations $\vec{\sigma}_b$ can be perfectly sampled from the MPS, with probability $\pi_B(\vec{\sigma}_b)$ and a cost $dD^3|B|$ per sample, where $d=2$ is the physical dimension and $D$ is the bond dimension. This gives direct access to both $\vec{\sigma}_b$ and $\pi_B(\vec{\sigma}_b)$. The conditioning of $\pi(t)$ on the outcome $\vec{\sigma}_B$ then allows for efficient computation of the MI between $A$ and $C$: for a given realisation $\vec{\sigma}_b = (\hat \sigma_1,\ldots,\hat\sigma_{|B|})$ we first form the $D\times D$-dimensional matrix 
\begin{equation}
F(\hat\sigma_1,\ldots,\hat\sigma_{|B|}) = \frac{X^{\hat\sigma_1}X^{\hat\sigma_2}\ldots X^{\hat\sigma_{|B|}}}{\pi(\hat\sigma_1,\hat\sigma_2,\ldots \hat\sigma_{|B|})}.
\end{equation}
Then, after forming the classical transfer matrix $T_{ab} = \sum_{\sigma}X_{ab}^\sigma$, we compute the leading left $\bra{V_L}$ and right $\ket{V_R}$ eigenvectors. Finally, we construct the marginal distribution on $A,C$, conditioned on the outcome $\vec{\sigma}_b$:
\begin{equation}
    \pi(\sigma_a,\sigma_c|\hat{\sigma}_1,\ldots\hat\sigma_{|B|}) = \bra{A^{\sigma_a}}F(\hat{\sigma}_1,\ldots\hat\sigma_{{|B|}})|C^{\sigma_c}\rangle.
\end{equation}
Here we assumed that $A$ and $C$ each consist of a single spin (the extension to larger subsystems is straightforward). The matrix has a small dimension $d\times d$ and the MI can be explicitly computed from the standard formula. Further details are given in \cite{SM}. Note that to exactly compute the CMI we still need a number of samples growing as $2^{|B|}$: however, as long as the average in Eq.~(\ref{eq:CMIsum}) is dominated by typical configurations of the $B$ subsystem, our method converges to a good estimator of the CMI within a reasonable number of samples. 
    
\emph{Markov length under Glauber dynamics.---} 
Before turning to the CMI, we first analyse the evolution of a more familiar quantity, the time-dependent spin-spin correlation function, $C_r(t) = \langle \sigma_0\sigma_r\rangle_{\pi(t)}$. This serves both to develop our intuition for the thermal noise case, and as a useful benchmark on our MPS simulations: the exact solution is derived in \cite{SM}, which gives us a concrete check on the accuracy of our simulations. We confirm in \cite{SM} that the MPS simulations correctly capture the autocorrelation function. 

We display the evolution of $C_r(t)$ in the inset of Fig.~\ref{fig:glauber}(b), for the case of ground state heating to a final temperature $\beta_f=1.5$. Darker purple lines indicate later time values, with a separation $t=150$. The initial and final thermal correlation lengths are marked with the blue and red lines respectively. The correlation function exhibits a clear two-slope profile,
\begin{equation}\label{twoshape}
    C_r(t) \approx e^{-{r/\xi_{\beta}}} +e^{-\gamma t-r/\xi_{\beta_i}},
\end{equation} 
where the thermal length is $\xi_\beta = -1/\log\tanh\beta$ and $\gamma$ is a rate depending on the initial and final temperatures. (Starting from the ground state $\beta_i=\infty$, the initial correlation length is infinite.) The correlation function is `thermalised' on an expanding scale $r^* = v^*t$ with a velocity $v^*=\gamma\frac{ \xi_{\beta_i}\xi_{\beta_f}}{\xi_{\beta_i}-\xi_{\beta_f}}$, while beyond this scale the remnants of initial correlations still dominate. Intuitively, the Gibbs state of a strictly local Hamiltonian cannot generate correlations with two distinct decay scales, so (by the Hammersley-Clifford theorem) the CMI should not be strictly short-range. We will now turn directly to the CMI and confirm this intuition.

First, we consider the case of the ground state heated to finite temperature. The information-theoretic argument for this case is closely analogous to that for quenching to infinite temperature: the Markov length should grow exponentially in time as the local magnetization decays exponentially, and it takes an increasingly large region $B$ to reliably infer the magnetization from majority vote. In Fig.~\ref{fig:glauber}(b), we confirm this prediction, by showing the space-time evolution of the CMI under Glauber dynamics with a final temperature $\beta_f = 1.5$. At late times, the CMI does not feature any appreciable decay for the sizes of the subsystem $B$ we study ($|B| \leq 200$), confirming the divergence of the Markov length as in the infinite temperature case. We show additional data in \cite{SM} showing that the Markov length grows exponentially in time, with a Markov length $\xi_M(t) \sim \xi_{\beta_f}(t)/m^2(t)$. This follows essentially from the arguments for ground state depolarization, with a renormalised block spin of size $\sim\xi_{\beta_f}$ entering the central limit theorem.

Interestingly, the spatial dependence of the CMI is not monotonic for finite final temperatures. The position of the CMI maximum, which we plot in Fig.~\ref{fig:glauber}(b) as the black dashed line, appears to travel ballistically in time; we interpret this initial growth of the CMI based on the picture of spin-spin correlations presented above. At short distances, $|B| \ll \mathcal{O}(v^*t)$, the dominant correlations between $A$ and $C$ are thermal correlations mediated by $B$, so we expect CMI to be small. Note that for the case of infinite temperature noise, $v^*= 0$, consistent with the monotonic decay of CMI in Eq.~(\ref{eq:collapse}). 

We finally consider the most general case, where the initial state is prepared at a temperature $\beta_i$, and subjected to Glauber noise at a higher temperature $\beta_f < \beta_i$. The initial state is now only correlated on a lengthscale $\xi_{\beta_i}$, which is finite for all non-zero temperatures in the 1d Ising model. For this reason, we do not expect the Markov length to diverge at any point along the dynamics (including at infinite time); indeed, the Markov length is bounded at all times by Theorem 2 in Ref.~\cite{zhang2025conditional}. Nevertheless, the Markov length is discontinuous in the late-time limit: as we argue below, the late-time Markov length is fixed by the initial thermal correlation length as $\xi_M = \xi_{\beta_i}/2$, and differs from that of the steady state, which has $\xi_M = 0$ by the Hammersley-Clifford theorem. We argue for this by considering the modified Glauber noise channel $ \Phi^B_\beta$ which acts only on the subsystem $B$. The late-time Markov length of the original and the modified channel is expected to be essentially the same, since it relates to the scaling of the CMI with respect to the size of $B$. Then, in the infinite time limit, the marginal distribution on $ABC$ is expected to approximately decouple according to $\pi({ABC}) \approx \pi({AC}) \theta_{\beta_f}(B)$, where $\theta_{\beta_f}(B)$ denotes the thermal distribution on $B$. We ignore the subtlety associated with the boundary conditions between $B$ and $AC$, which should not be important for the Markov length behavior as long as $\beta_f < \beta_i$. Due to the independence of subsystems $AC$ and $B$, $I_{\pi(\infty)}(A:C|B) \approx I_{\pi(0)}(A:C)$, in terms of the initial MI. The mutual information for the initial thermal state can be calculated exactly~\cite{SM}; it scales as the square of the spin-spin correlation function, i.e., $I_{\pi(0)}(A:C) \sim \exp(-2|B|/\xi_{\beta_i})$. From this we immediately read off that $\xi_M = \xi_{\beta_i}/2$.

We test our prediction in Fig.~\ref{fig:glauber}(c), where we display the space-time behaviour of the CMI for initial temperature $\beta_i = 2$ and final temperature $\beta_f = 1.4$. At late times, the CMI decays exponentially in the distance $|B|$, with a finite Markov length. In the inset, we show the scaling of the late-time Markov length as a function of the initial temperature $\beta_i$. The black points show numerically extracted values while the dashed line is the curve $\xi_M = \xi_{\beta_i}/2$. The match to the data is almost exact, confirming the argument given above. Our results are consistent with the divergence of the Markov length starting from the ground state, since in this case the thermal lengthscale diverges. In the Supplemental Material, we furthermore relate the growth of the Markov length to the \emph{Lyapunov spectrum} of the random matrix product $F(\vec{\sigma}_b)$: we find that the Markov length is well predicted by the leading values of the Lyapunov spectrum for large $|B|$.

\emph{Discussion.---} We have presented evidence that the Markov length grows exponentially in time when an ordered state thermalizes into the disordered phase. Thus, even though the high-temperature dynamics mixes rapidly for conventional observables, information-theoretic quantities can exhibit divergent length-scales, and remain far from their steady-state values at any finite time. This divergent length-scale has the physical interpretation that the state of a system as it heats up is increasingly non-Gibbsian: any parent Hamiltonian for which this state is a Gibbs state must have a range that diverges exponentially in time. We have provided an information-theoretic interpretation of this diverging length-scale in terms of an error correction threshold that (in this simple model) occurs at $t = \infty$. This argument would further suggest that if our initial state had been a code with a lower threshold, $\xi_M$ would have exhibited a finite-time divergence: such divergences are associated with transitions in classical inference~\cite{zdeborova2016statistical, PhysRevLett.129.200602, nahum2025bayesian, gopalakrishnan2025monitored, kim2025measurement, putz2025learning}. For the one-dimensional geometries that allow MPS representations, we have not found examples of this behavior; finding efficient ways to compute CMI in higher-dimensional models (e.g., using variational methods from machine learning~\cite{alemi2016deep}) is an interesting task for future studies. An especially interesting case is that of the four-dimensional toric code, in which one can leave the coding phase either through a mixed-state transition driven by noise (at which the Markov length must diverge) or an equilibrium thermal transition (across which the Markov length remains exactly zero): it would be interesting to understand how these transitions fit together in a global phase diagram.

In the classical setting, the Hammersley-Clifford theorem is a powerful and convenient diagnostic for non-Gibbsian states. It was shown very recently that quantum Gibbs states obey a Markov-like property~\cite{chen2025quantum} (albeit with a Markov length that diverges at low temperature). Thus, mixed states with long-range CMI cannot have local parent Hamiltonians with nondegenerate spectra. Establishing whether such non-Gibbsian states form stable mixed-state phases, and when they arise from local dissipative evolution, are important open questions.

\begin{acknowledgments}
S.G. thanks Tim Hsieh, David Huse, Curt von Keyserlingk, Benedikt Placke, Tibor Rakovszky, and especially Yifan Zhang for helpful discussions and collaborations on related topics. S.G. was supported by NSF QuSEC-TAQS OSI 232676.
\end{acknowledgments}

\acknowledgments
\bibliography{bib.bib}

\begin{thebibliography}{42}%
\makeatletter
\providecommand \@ifxundefined [1]{%
 \@ifx{#1\undefined}
}%
\providecommand \@ifnum [1]{%
 \ifnum #1\expandafter \@firstoftwo
 \else \expandafter \@secondoftwo
 \fi
}%
\providecommand \@ifx [1]{%
 \ifx #1\expandafter \@firstoftwo
 \else \expandafter \@secondoftwo
 \fi
}%
\providecommand \natexlab [1]{#1}%
\providecommand \enquote  [1]{``#1''}%
\providecommand \bibnamefont  [1]{#1}%
\providecommand \bibfnamefont [1]{#1}%
\providecommand \citenamefont [1]{#1}%
\providecommand \href@noop [0]{\@secondoftwo}%
\providecommand \href [0]{\begingroup \@sanitize@url \@href}%
\providecommand \@href[1]{\@@startlink{#1}\@@href}%
\providecommand \@@href[1]{\endgroup#1\@@endlink}%
\providecommand \@sanitize@url [0]{\catcode `\\12\catcode `\$12\catcode `\&12\catcode `\#12\catcode `\^12\catcode `\_12\catcode `\%12\relax}%
\providecommand \@@startlink[1]{}%
\providecommand \@@endlink[0]{}%
\providecommand \url  [0]{\begingroup\@sanitize@url \@url }%
\providecommand \@url [1]{\endgroup\@href {#1}{\urlprefix }}%
\providecommand \urlprefix  [0]{URL }%
\providecommand \Eprint [0]{\href }%
\providecommand \doibase [0]{https://doi.org/}%
\providecommand \selectlanguage [0]{\@gobble}%
\providecommand \bibinfo  [0]{\@secondoftwo}%
\providecommand \bibfield  [0]{\@secondoftwo}%
\providecommand \translation [1]{[#1]}%
\providecommand \BibitemOpen [0]{}%
\providecommand \bibitemStop [0]{}%
\providecommand \bibitemNoStop [0]{.\EOS\space}%
\providecommand \EOS [0]{\spacefactor3000\relax}%
\providecommand \BibitemShut  [1]{\csname bibitem#1\endcsname}%
\let\auto@bib@innerbib\@empty
\bibitem [{\citenamefont {Dennis}\ \emph {et~al.}(2002)\citenamefont {Dennis}, \citenamefont {Kitaev}, \citenamefont {Landahl},\ and\ \citenamefont {Preskill}}]{dennis2002topological}%
  \BibitemOpen
  \bibfield  {author} {\bibinfo {author} {\bibfnamefont {E.}~\bibnamefont {Dennis}}, \bibinfo {author} {\bibfnamefont {A.}~\bibnamefont {Kitaev}}, \bibinfo {author} {\bibfnamefont {A.}~\bibnamefont {Landahl}},\ and\ \bibinfo {author} {\bibfnamefont {J.}~\bibnamefont {Preskill}},\ }\bibfield  {title} {\bibinfo {title} {Topological quantum memory},\ }\href@noop {} {\bibfield  {journal} {\bibinfo  {journal} {Journal of Mathematical Physics}\ }\textbf {\bibinfo {volume} {43}},\ \bibinfo {pages} {4452} (\bibinfo {year} {2002})}\BibitemShut {NoStop}%
\bibitem [{\citenamefont {Fan}\ \emph {et~al.}(2024)\citenamefont {Fan}, \citenamefont {Bao}, \citenamefont {Altman},\ and\ \citenamefont {Vishwanath}}]{PRXQuantum.5.020343}%
  \BibitemOpen
  \bibfield  {author} {\bibinfo {author} {\bibfnamefont {R.}~\bibnamefont {Fan}}, \bibinfo {author} {\bibfnamefont {Y.}~\bibnamefont {Bao}}, \bibinfo {author} {\bibfnamefont {E.}~\bibnamefont {Altman}},\ and\ \bibinfo {author} {\bibfnamefont {A.}~\bibnamefont {Vishwanath}},\ }\bibfield  {title} {\bibinfo {title} {Diagnostics of mixed-state topological order and breakdown of quantum memory},\ }\href {https://doi.org/10.1103/PRXQuantum.5.020343} {\bibfield  {journal} {\bibinfo  {journal} {PRX Quantum}\ }\textbf {\bibinfo {volume} {5}},\ \bibinfo {pages} {020343} (\bibinfo {year} {2024})}\BibitemShut {NoStop}%
\bibitem [{\citenamefont {Su}\ \emph {et~al.}(2024)\citenamefont {Su}, \citenamefont {Yang},\ and\ \citenamefont {Jian}}]{PhysRevB.110.085158}%
  \BibitemOpen
  \bibfield  {author} {\bibinfo {author} {\bibfnamefont {K.}~\bibnamefont {Su}}, \bibinfo {author} {\bibfnamefont {Z.}~\bibnamefont {Yang}},\ and\ \bibinfo {author} {\bibfnamefont {C.-M.}\ \bibnamefont {Jian}},\ }\bibfield  {title} {\bibinfo {title} {Tapestry of dualities in decohered quantum error correction codes},\ }\href {https://doi.org/10.1103/PhysRevB.110.085158} {\bibfield  {journal} {\bibinfo  {journal} {Phys. Rev. B}\ }\textbf {\bibinfo {volume} {110}},\ \bibinfo {pages} {085158} (\bibinfo {year} {2024})}\BibitemShut {NoStop}%
\bibitem [{\citenamefont {Lessa}\ \emph {et~al.}(2025)\citenamefont {Lessa}, \citenamefont {Cheng},\ and\ \citenamefont {Wang}}]{PhysRevX.15.011069}%
  \BibitemOpen
  \bibfield  {author} {\bibinfo {author} {\bibfnamefont {L.~A.}\ \bibnamefont {Lessa}}, \bibinfo {author} {\bibfnamefont {M.}~\bibnamefont {Cheng}},\ and\ \bibinfo {author} {\bibfnamefont {C.}~\bibnamefont {Wang}},\ }\bibfield  {title} {\bibinfo {title} {Mixed-state quantum anomaly and multipartite entanglement},\ }\href {https://doi.org/10.1103/PhysRevX.15.011069} {\bibfield  {journal} {\bibinfo  {journal} {Phys. Rev. X}\ }\textbf {\bibinfo {volume} {15}},\ \bibinfo {pages} {011069} (\bibinfo {year} {2025})}\BibitemShut {NoStop}%
\bibitem [{\citenamefont {Coser}\ and\ \citenamefont {P{\'e}rez-Garc{\'\i}a}(2019)}]{coser2019classification}%
  \BibitemOpen
  \bibfield  {author} {\bibinfo {author} {\bibfnamefont {A.}~\bibnamefont {Coser}}\ and\ \bibinfo {author} {\bibfnamefont {D.}~\bibnamefont {P{\'e}rez-Garc{\'\i}a}},\ }\bibfield  {title} {\bibinfo {title} {Classification of phases for mixed states via fast dissipative evolution},\ }\href@noop {} {\bibfield  {journal} {\bibinfo  {journal} {Quantum}\ }\textbf {\bibinfo {volume} {3}},\ \bibinfo {pages} {174} (\bibinfo {year} {2019})}\BibitemShut {NoStop}%
\bibitem [{\citenamefont {Poulin}(2010)}]{PhysRevLett.104.190401}%
  \BibitemOpen
  \bibfield  {author} {\bibinfo {author} {\bibfnamefont {D.}~\bibnamefont {Poulin}},\ }\bibfield  {title} {\bibinfo {title} {Lieb-robinson bound and locality for general markovian quantum dynamics},\ }\href {https://doi.org/10.1103/PhysRevLett.104.190401} {\bibfield  {journal} {\bibinfo  {journal} {Phys. Rev. Lett.}\ }\textbf {\bibinfo {volume} {104}},\ \bibinfo {pages} {190401} (\bibinfo {year} {2010})}\BibitemShut {NoStop}%
\bibitem [{\citenamefont {Rakovszky}\ \emph {et~al.}(2024)\citenamefont {Rakovszky}, \citenamefont {Gopalakrishnan},\ and\ \citenamefont {von Keyserlingk}}]{PhysRevX.14.041031}%
  \BibitemOpen
  \bibfield  {author} {\bibinfo {author} {\bibfnamefont {T.}~\bibnamefont {Rakovszky}}, \bibinfo {author} {\bibfnamefont {S.}~\bibnamefont {Gopalakrishnan}},\ and\ \bibinfo {author} {\bibfnamefont {C.}~\bibnamefont {von Keyserlingk}},\ }\bibfield  {title} {\bibinfo {title} {Defining stable phases of open quantum systems},\ }\href {https://doi.org/10.1103/PhysRevX.14.041031} {\bibfield  {journal} {\bibinfo  {journal} {Phys. Rev. X}\ }\textbf {\bibinfo {volume} {14}},\ \bibinfo {pages} {041031} (\bibinfo {year} {2024})}\BibitemShut {NoStop}%
\bibitem [{\citenamefont {Sang}\ and\ \citenamefont {Hsieh}(2025)}]{shengqi}%
  \BibitemOpen
  \bibfield  {author} {\bibinfo {author} {\bibfnamefont {S.}~\bibnamefont {Sang}}\ and\ \bibinfo {author} {\bibfnamefont {T.~H.}\ \bibnamefont {Hsieh}},\ }\bibfield  {title} {\bibinfo {title} {Stability of mixed-state quantum phases via finite markov length},\ }\href {https://doi.org/10.1103/PhysRevLett.134.070403} {\bibfield  {journal} {\bibinfo  {journal} {Phys. Rev. Lett.}\ }\textbf {\bibinfo {volume} {134}},\ \bibinfo {pages} {070403} (\bibinfo {year} {2025})}\BibitemShut {NoStop}%
\bibitem [{\citenamefont {Sang}\ \emph {et~al.}(2023)\citenamefont {Sang}, \citenamefont {Li}, \citenamefont {Hsieh},\ and\ \citenamefont {Yoshida}}]{PRXQuantum.4.040332}%
  \BibitemOpen
  \bibfield  {author} {\bibinfo {author} {\bibfnamefont {S.}~\bibnamefont {Sang}}, \bibinfo {author} {\bibfnamefont {Z.}~\bibnamefont {Li}}, \bibinfo {author} {\bibfnamefont {T.~H.}\ \bibnamefont {Hsieh}},\ and\ \bibinfo {author} {\bibfnamefont {B.}~\bibnamefont {Yoshida}},\ }\bibfield  {title} {\bibinfo {title} {Ultrafast entanglement dynamics in monitored quantum circuits},\ }\href {https://doi.org/10.1103/PRXQuantum.4.040332} {\bibfield  {journal} {\bibinfo  {journal} {PRX Quantum}\ }\textbf {\bibinfo {volume} {4}},\ \bibinfo {pages} {040332} (\bibinfo {year} {2023})}\BibitemShut {NoStop}%
\bibitem [{\citenamefont {Zhang}\ and\ \citenamefont {Gopalakrishnan}(2024)}]{PhysRevA.110.032426}%
  \BibitemOpen
  \bibfield  {author} {\bibinfo {author} {\bibfnamefont {Y.}~\bibnamefont {Zhang}}\ and\ \bibinfo {author} {\bibfnamefont {S.}~\bibnamefont {Gopalakrishnan}},\ }\bibfield  {title} {\bibinfo {title} {Nonlocal growth of quantum conditional mutual information under decoherence},\ }\href {https://doi.org/10.1103/PhysRevA.110.032426} {\bibfield  {journal} {\bibinfo  {journal} {Phys. Rev. A}\ }\textbf {\bibinfo {volume} {110}},\ \bibinfo {pages} {032426} (\bibinfo {year} {2024})}\BibitemShut {NoStop}%
\bibitem [{\citenamefont {Lee}\ \emph {et~al.}(2024)\citenamefont {Lee}, \citenamefont {Oh}, \citenamefont {Wong}, \citenamefont {Chen},\ and\ \citenamefont {Jiang}}]{PhysRevLett.133.200402}%
  \BibitemOpen
  \bibfield  {author} {\bibinfo {author} {\bibfnamefont {S.-u.}\ \bibnamefont {Lee}}, \bibinfo {author} {\bibfnamefont {C.}~\bibnamefont {Oh}}, \bibinfo {author} {\bibfnamefont {Y.}~\bibnamefont {Wong}}, \bibinfo {author} {\bibfnamefont {S.}~\bibnamefont {Chen}},\ and\ \bibinfo {author} {\bibfnamefont {L.}~\bibnamefont {Jiang}},\ }\bibfield  {title} {\bibinfo {title} {Universal spreading of conditional mutual information in noisy random circuits},\ }\href {https://doi.org/10.1103/PhysRevLett.133.200402} {\bibfield  {journal} {\bibinfo  {journal} {Phys. Rev. Lett.}\ }\textbf {\bibinfo {volume} {133}},\ \bibinfo {pages} {200402} (\bibinfo {year} {2024})}\BibitemShut {NoStop}%
\bibitem [{\citenamefont {Bao}\ \emph {et~al.}(2024)\citenamefont {Bao}, \citenamefont {Block},\ and\ \citenamefont {Altman}}]{PhysRevLett.132.030401}%
  \BibitemOpen
  \bibfield  {author} {\bibinfo {author} {\bibfnamefont {Y.}~\bibnamefont {Bao}}, \bibinfo {author} {\bibfnamefont {M.}~\bibnamefont {Block}},\ and\ \bibinfo {author} {\bibfnamefont {E.}~\bibnamefont {Altman}},\ }\bibfield  {title} {\bibinfo {title} {Finite-time teleportation phase transition in random quantum circuits},\ }\href {https://doi.org/10.1103/PhysRevLett.132.030401} {\bibfield  {journal} {\bibinfo  {journal} {Phys. Rev. Lett.}\ }\textbf {\bibinfo {volume} {132}},\ \bibinfo {pages} {030401} (\bibinfo {year} {2024})}\BibitemShut {NoStop}%
\bibitem [{\citenamefont {Napp}\ \emph {et~al.}(2022)\citenamefont {Napp}, \citenamefont {La~Placa}, \citenamefont {Dalzell}, \citenamefont {Brand\~ao},\ and\ \citenamefont {Harrow}}]{PhysRevX.12.021021}%
  \BibitemOpen
  \bibfield  {author} {\bibinfo {author} {\bibfnamefont {J.~C.}\ \bibnamefont {Napp}}, \bibinfo {author} {\bibfnamefont {R.~L.}\ \bibnamefont {La~Placa}}, \bibinfo {author} {\bibfnamefont {A.~M.}\ \bibnamefont {Dalzell}}, \bibinfo {author} {\bibfnamefont {F.~G. S.~L.}\ \bibnamefont {Brand\~ao}},\ and\ \bibinfo {author} {\bibfnamefont {A.~W.}\ \bibnamefont {Harrow}},\ }\bibfield  {title} {\bibinfo {title} {Efficient classical simulation of random shallow 2d quantum circuits},\ }\href {https://doi.org/10.1103/PhysRevX.12.021021} {\bibfield  {journal} {\bibinfo  {journal} {Phys. Rev. X}\ }\textbf {\bibinfo {volume} {12}},\ \bibinfo {pages} {021021} (\bibinfo {year} {2022})}\BibitemShut {NoStop}%
\bibitem [{\citenamefont {Lu}\ \emph {et~al.}(2022)\citenamefont {Lu}, \citenamefont {Lessa}, \citenamefont {Kim},\ and\ \citenamefont {Hsieh}}]{PRXQuantum.3.040337}%
  \BibitemOpen
  \bibfield  {author} {\bibinfo {author} {\bibfnamefont {T.-C.}\ \bibnamefont {Lu}}, \bibinfo {author} {\bibfnamefont {L.~A.}\ \bibnamefont {Lessa}}, \bibinfo {author} {\bibfnamefont {I.~H.}\ \bibnamefont {Kim}},\ and\ \bibinfo {author} {\bibfnamefont {T.~H.}\ \bibnamefont {Hsieh}},\ }\bibfield  {title} {\bibinfo {title} {Measurement as a shortcut to long-range entangled quantum matter},\ }\href {https://doi.org/10.1103/PRXQuantum.3.040337} {\bibfield  {journal} {\bibinfo  {journal} {PRX Quantum}\ }\textbf {\bibinfo {volume} {3}},\ \bibinfo {pages} {040337} (\bibinfo {year} {2022})}\BibitemShut {NoStop}%
\bibitem [{\citenamefont {Sang}\ \emph {et~al.}(2025)\citenamefont {Sang}, \citenamefont {Lessa}, \citenamefont {Mong}, \citenamefont {Grover}, \citenamefont {Wang},\ and\ \citenamefont {Hsieh}}]{sang2025mixed}%
  \BibitemOpen
  \bibfield  {author} {\bibinfo {author} {\bibfnamefont {S.}~\bibnamefont {Sang}}, \bibinfo {author} {\bibfnamefont {L.~A.}\ \bibnamefont {Lessa}}, \bibinfo {author} {\bibfnamefont {R.~S.}\ \bibnamefont {Mong}}, \bibinfo {author} {\bibfnamefont {T.}~\bibnamefont {Grover}}, \bibinfo {author} {\bibfnamefont {C.}~\bibnamefont {Wang}},\ and\ \bibinfo {author} {\bibfnamefont {T.~H.}\ \bibnamefont {Hsieh}},\ }\bibfield  {title} {\bibinfo {title} {Mixed-state phases from local reversibility},\ }\href@noop {} {\bibfield  {journal} {\bibinfo  {journal} {arXiv preprint arXiv:2507.02292}\ } (\bibinfo {year} {2025})}\BibitemShut {NoStop}%
\bibitem [{\citenamefont {Hammersley}\ and\ \citenamefont {Clifford}(1971)}]{hammersley1971markov}%
  \BibitemOpen
  \bibfield  {author} {\bibinfo {author} {\bibfnamefont {J.~M.}\ \bibnamefont {Hammersley}}\ and\ \bibinfo {author} {\bibfnamefont {P.}~\bibnamefont {Clifford}},\ }\bibfield  {title} {\bibinfo {title} {Markov fields on finite graphs and lattices},\ }\href@noop {} {\bibfield  {journal} {\bibinfo  {journal} {Unpublished manuscript}\ }\textbf {\bibinfo {volume} {46}} (\bibinfo {year} {1971})}\BibitemShut {NoStop}%
\bibitem [{\citenamefont {Zhang}\ and\ \citenamefont {Gopalakrishnan}(2025)}]{zhang2025conditional}%
  \BibitemOpen
  \bibfield  {author} {\bibinfo {author} {\bibfnamefont {Y.}~\bibnamefont {Zhang}}\ and\ \bibinfo {author} {\bibfnamefont {S.}~\bibnamefont {Gopalakrishnan}},\ }\bibfield  {title} {\bibinfo {title} {Conditional mutual information and information-theoretic phases of decohered gibbs states},\ }\href@noop {} {\bibfield  {journal} {\bibinfo  {journal} {arXiv preprint arXiv:2502.13210}\ } (\bibinfo {year} {2025})}\BibitemShut {NoStop}%
\bibitem [{SM()}]{SM}%
  \BibitemOpen
  \href@noop {} {\bibinfo  {journal} {See Supplementary Material}\ }\BibitemShut {NoStop}%
\bibitem [{\citenamefont {Glauber}(1963)}]{glauber1963time}%
  \BibitemOpen
\bibfield  {journal} {  }\bibfield  {author} {\bibinfo {author} {\bibfnamefont {R.~J.}\ \bibnamefont {Glauber}},\ }\bibfield  {title} {\bibinfo {title} {Time-dependent statistics of the ising model},\ }\href@noop {} {\bibfield  {journal} {\bibinfo  {journal} {Journal of mathematical physics}\ }\textbf {\bibinfo {volume} {4}},\ \bibinfo {pages} {294} (\bibinfo {year} {1963})}\BibitemShut {NoStop}%
\bibitem [{\citenamefont {Johnson}\ \emph {et~al.}(2010)\citenamefont {Johnson}, \citenamefont {Clark},\ and\ \citenamefont {Jaksch}}]{johnson2010dynamical}%
  \BibitemOpen
  \bibfield  {author} {\bibinfo {author} {\bibfnamefont {T.}~\bibnamefont {Johnson}}, \bibinfo {author} {\bibfnamefont {S.}~\bibnamefont {Clark}},\ and\ \bibinfo {author} {\bibfnamefont {D.}~\bibnamefont {Jaksch}},\ }\bibfield  {title} {\bibinfo {title} {Dynamical simulations of classical stochastic systems using matrix product states},\ }\href@noop {} {\bibfield  {journal} {\bibinfo  {journal} {Physical Review E—Statistical, Nonlinear, and Soft Matter Physics}\ }\textbf {\bibinfo {volume} {82}},\ \bibinfo {pages} {036702} (\bibinfo {year} {2010})}\BibitemShut {NoStop}%
\bibitem [{\citenamefont {Johnson}\ \emph {et~al.}(2015)\citenamefont {Johnson}, \citenamefont {Elliott}, \citenamefont {Clark},\ and\ \citenamefont {Jaksch}}]{johnson2015capturing}%
  \BibitemOpen
  \bibfield  {author} {\bibinfo {author} {\bibfnamefont {T.}~\bibnamefont {Johnson}}, \bibinfo {author} {\bibfnamefont {T.}~\bibnamefont {Elliott}}, \bibinfo {author} {\bibfnamefont {S.}~\bibnamefont {Clark}},\ and\ \bibinfo {author} {\bibfnamefont {D.}~\bibnamefont {Jaksch}},\ }\bibfield  {title} {\bibinfo {title} {Capturing exponential variance using polynomial resources: applying tensor networks to nonequilibrium stochastic processes},\ }\href@noop {} {\bibfield  {journal} {\bibinfo  {journal} {Physical review letters}\ }\textbf {\bibinfo {volume} {114}},\ \bibinfo {pages} {090602} (\bibinfo {year} {2015})}\BibitemShut {NoStop}%
\bibitem [{\citenamefont {Banuls}\ and\ \citenamefont {Garrahan}(2019)}]{banuls2019using}%
  \BibitemOpen
  \bibfield  {author} {\bibinfo {author} {\bibfnamefont {M.~C.}\ \bibnamefont {Banuls}}\ and\ \bibinfo {author} {\bibfnamefont {J.~P.}\ \bibnamefont {Garrahan}},\ }\bibfield  {title} {\bibinfo {title} {Using matrix product states to study the dynamical large deviations of kinetically constrained models},\ }\href@noop {} {\bibfield  {journal} {\bibinfo  {journal} {Physical review letters}\ }\textbf {\bibinfo {volume} {123}},\ \bibinfo {pages} {200601} (\bibinfo {year} {2019})}\BibitemShut {NoStop}%
\bibitem [{\citenamefont {Causer}\ \emph {et~al.}(2022)\citenamefont {Causer}, \citenamefont {Ba{\~n}uls},\ and\ \citenamefont {Garrahan}}]{causer2022finite}%
  \BibitemOpen
  \bibfield  {author} {\bibinfo {author} {\bibfnamefont {L.}~\bibnamefont {Causer}}, \bibinfo {author} {\bibfnamefont {M.~C.}\ \bibnamefont {Ba{\~n}uls}},\ and\ \bibinfo {author} {\bibfnamefont {J.~P.}\ \bibnamefont {Garrahan}},\ }\bibfield  {title} {\bibinfo {title} {Finite time large deviations via matrix product states},\ }\href@noop {} {\bibfield  {journal} {\bibinfo  {journal} {Physical Review Letters}\ }\textbf {\bibinfo {volume} {128}},\ \bibinfo {pages} {090605} (\bibinfo {year} {2022})}\BibitemShut {NoStop}%
\bibitem [{\citenamefont {Gu}\ and\ \citenamefont {Zhang}(2022)}]{gu2022tensor}%
  \BibitemOpen
  \bibfield  {author} {\bibinfo {author} {\bibfnamefont {J.}~\bibnamefont {Gu}}\ and\ \bibinfo {author} {\bibfnamefont {F.}~\bibnamefont {Zhang}},\ }\bibfield  {title} {\bibinfo {title} {Tensor-network approaches to counting statistics for the current in a boundary-driven diffusive system},\ }\href@noop {} {\bibfield  {journal} {\bibinfo  {journal} {New Journal of Physics}\ }\textbf {\bibinfo {volume} {24}},\ \bibinfo {pages} {113022} (\bibinfo {year} {2022})}\BibitemShut {NoStop}%
\bibitem [{\citenamefont {Zdeborov{\'a}}\ and\ \citenamefont {Krzakala}(2016)}]{zdeborova2016statistical}%
  \BibitemOpen
  \bibfield  {author} {\bibinfo {author} {\bibfnamefont {L.}~\bibnamefont {Zdeborov{\'a}}}\ and\ \bibinfo {author} {\bibfnamefont {F.}~\bibnamefont {Krzakala}},\ }\bibfield  {title} {\bibinfo {title} {Statistical physics of inference: Thresholds and algorithms},\ }\href@noop {} {\bibfield  {journal} {\bibinfo  {journal} {Advances in Physics}\ }\textbf {\bibinfo {volume} {65}},\ \bibinfo {pages} {453} (\bibinfo {year} {2016})}\BibitemShut {NoStop}%
\bibitem [{\citenamefont {Barratt}\ \emph {et~al.}(2022)\citenamefont {Barratt}, \citenamefont {Agrawal}, \citenamefont {Potter}, \citenamefont {Gopalakrishnan},\ and\ \citenamefont {Vasseur}}]{PhysRevLett.129.200602}%
  \BibitemOpen
  \bibfield  {author} {\bibinfo {author} {\bibfnamefont {F.}~\bibnamefont {Barratt}}, \bibinfo {author} {\bibfnamefont {U.}~\bibnamefont {Agrawal}}, \bibinfo {author} {\bibfnamefont {A.~C.}\ \bibnamefont {Potter}}, \bibinfo {author} {\bibfnamefont {S.}~\bibnamefont {Gopalakrishnan}},\ and\ \bibinfo {author} {\bibfnamefont {R.}~\bibnamefont {Vasseur}},\ }\bibfield  {title} {\bibinfo {title} {Transitions in the learnability of global charges from local measurements},\ }\href {https://doi.org/10.1103/PhysRevLett.129.200602} {\bibfield  {journal} {\bibinfo  {journal} {Phys. Rev. Lett.}\ }\textbf {\bibinfo {volume} {129}},\ \bibinfo {pages} {200602} (\bibinfo {year} {2022})}\BibitemShut {NoStop}%
\bibitem [{\citenamefont {Nahum}\ and\ \citenamefont {Jacobsen}(2025)}]{nahum2025bayesian}%
  \BibitemOpen
  \bibfield  {author} {\bibinfo {author} {\bibfnamefont {A.}~\bibnamefont {Nahum}}\ and\ \bibinfo {author} {\bibfnamefont {J.~L.}\ \bibnamefont {Jacobsen}},\ }\bibfield  {title} {\bibinfo {title} {Bayesian critical points in classical lattice models},\ }\href@noop {} {\bibfield  {journal} {\bibinfo  {journal} {arXiv preprint arXiv:2504.01264}\ } (\bibinfo {year} {2025})}\BibitemShut {NoStop}%
\bibitem [{\citenamefont {Gopalakrishnan}\ \emph {et~al.}(2025)\citenamefont {Gopalakrishnan}, \citenamefont {McCulloch},\ and\ \citenamefont {Vasseur}}]{gopalakrishnan2025monitored}%
  \BibitemOpen
  \bibfield  {author} {\bibinfo {author} {\bibfnamefont {S.}~\bibnamefont {Gopalakrishnan}}, \bibinfo {author} {\bibfnamefont {E.}~\bibnamefont {McCulloch}},\ and\ \bibinfo {author} {\bibfnamefont {R.}~\bibnamefont {Vasseur}},\ }\bibfield  {title} {\bibinfo {title} {Monitored fluctuating hydrodynamics},\ }\href@noop {} {\bibfield  {journal} {\bibinfo  {journal} {arXiv preprint arXiv:2504.02734}\ } (\bibinfo {year} {2025})}\BibitemShut {NoStop}%
\bibitem [{\citenamefont {Kim}\ \emph {et~al.}(2025)\citenamefont {Kim}, \citenamefont {von Keyserlingk},\ and\ \citenamefont {Lamacraft}}]{kim2025measurement}%
  \BibitemOpen
  \bibfield  {author} {\bibinfo {author} {\bibfnamefont {S.~W.~P.}\ \bibnamefont {Kim}}, \bibinfo {author} {\bibfnamefont {C.}~\bibnamefont {von Keyserlingk}},\ and\ \bibinfo {author} {\bibfnamefont {A.}~\bibnamefont {Lamacraft}},\ }\bibfield  {title} {\bibinfo {title} {Measurement-induced phase transitions in quantum inference problems and quantum hidden markov models},\ }\href@noop {} {\bibfield  {journal} {\bibinfo  {journal} {arXiv preprint arXiv:2504.08888}\ } (\bibinfo {year} {2025})}\BibitemShut {NoStop}%
\bibitem [{\citenamefont {P{\"u}tz}\ \emph {et~al.}(2025)\citenamefont {P{\"u}tz}, \citenamefont {Garratt}, \citenamefont {Nishimori}, \citenamefont {Trebst},\ and\ \citenamefont {Zhu}}]{putz2025learning}%
  \BibitemOpen
  \bibfield  {author} {\bibinfo {author} {\bibfnamefont {M.}~\bibnamefont {P{\"u}tz}}, \bibinfo {author} {\bibfnamefont {S.~J.}\ \bibnamefont {Garratt}}, \bibinfo {author} {\bibfnamefont {H.}~\bibnamefont {Nishimori}}, \bibinfo {author} {\bibfnamefont {S.}~\bibnamefont {Trebst}},\ and\ \bibinfo {author} {\bibfnamefont {G.-Y.}\ \bibnamefont {Zhu}},\ }\bibfield  {title} {\bibinfo {title} {Learning transitions in classical ising models and deformed toric codes},\ }\href@noop {} {\bibfield  {journal} {\bibinfo  {journal} {arXiv preprint arXiv:2504.12385}\ } (\bibinfo {year} {2025})}\BibitemShut {NoStop}%
\bibitem [{\citenamefont {Alemi}\ \emph {et~al.}(2016)\citenamefont {Alemi}, \citenamefont {Fischer}, \citenamefont {Dillon},\ and\ \citenamefont {Murphy}}]{alemi2016deep}%
  \BibitemOpen
  \bibfield  {author} {\bibinfo {author} {\bibfnamefont {A.~A.}\ \bibnamefont {Alemi}}, \bibinfo {author} {\bibfnamefont {I.}~\bibnamefont {Fischer}}, \bibinfo {author} {\bibfnamefont {J.~V.}\ \bibnamefont {Dillon}},\ and\ \bibinfo {author} {\bibfnamefont {K.}~\bibnamefont {Murphy}},\ }\bibfield  {title} {\bibinfo {title} {Deep variational information bottleneck},\ }\href@noop {} {\bibfield  {journal} {\bibinfo  {journal} {arXiv preprint arXiv:1612.00410}\ } (\bibinfo {year} {2016})}\BibitemShut {NoStop}%
\bibitem [{\citenamefont {Chen}\ and\ \citenamefont {Rouz{\'e}}(2025)}]{chen2025quantum}%
  \BibitemOpen
  \bibfield  {author} {\bibinfo {author} {\bibfnamefont {C.-F.}\ \bibnamefont {Chen}}\ and\ \bibinfo {author} {\bibfnamefont {C.}~\bibnamefont {Rouz{\'e}}},\ }\bibfield  {title} {\bibinfo {title} {Quantum gibbs states are locally markovian},\ }\href@noop {} {\bibfield  {journal} {\bibinfo  {journal} {arXiv preprint arXiv:2504.02208}\ } (\bibinfo {year} {2025})}\BibitemShut {NoStop}%
\bibitem [{\citenamefont {Vidal}(2004)}]{vidal2004efficient}%
  \BibitemOpen
  \bibfield  {author} {\bibinfo {author} {\bibfnamefont {G.}~\bibnamefont {Vidal}},\ }\bibfield  {title} {\bibinfo {title} {Efficient simulation of one-dimensional quantum many-body systems},\ }\href@noop {} {\bibfield  {journal} {\bibinfo  {journal} {Physical review letters}\ }\textbf {\bibinfo {volume} {93}},\ \bibinfo {pages} {040502} (\bibinfo {year} {2004})}\BibitemShut {NoStop}%
\bibitem [{\citenamefont {Verstraete}\ \emph {et~al.}(2004)\citenamefont {Verstraete}, \citenamefont {Garcia-Ripoll},\ and\ \citenamefont {Cirac}}]{verstraete2004matrix}%
  \BibitemOpen
  \bibfield  {author} {\bibinfo {author} {\bibfnamefont {F.}~\bibnamefont {Verstraete}}, \bibinfo {author} {\bibfnamefont {J.~J.}\ \bibnamefont {Garcia-Ripoll}},\ and\ \bibinfo {author} {\bibfnamefont {J.~I.}\ \bibnamefont {Cirac}},\ }\bibfield  {title} {\bibinfo {title} {Matrix product density operators: Simulation of finite-temperature and dissipative systems},\ }\href@noop {} {\bibfield  {journal} {\bibinfo  {journal} {Physical review letters}\ }\textbf {\bibinfo {volume} {93}},\ \bibinfo {pages} {207204} (\bibinfo {year} {2004})}\BibitemShut {NoStop}%
\bibitem [{\citenamefont {Orus}\ and\ \citenamefont {Vidal}(2008)}]{orus2008infinite}%
  \BibitemOpen
  \bibfield  {author} {\bibinfo {author} {\bibfnamefont {R.}~\bibnamefont {Orus}}\ and\ \bibinfo {author} {\bibfnamefont {G.}~\bibnamefont {Vidal}},\ }\bibfield  {title} {\bibinfo {title} {Infinite time-evolving block decimation algorithm beyond unitary evolution},\ }\href@noop {} {\bibfield  {journal} {\bibinfo  {journal} {Physical Review B—Condensed Matter and Materials Physics}\ }\textbf {\bibinfo {volume} {78}},\ \bibinfo {pages} {155117} (\bibinfo {year} {2008})}\BibitemShut {NoStop}%
\bibitem [{\citenamefont {Vanderstraeten}\ \emph {et~al.}(2019)\citenamefont {Vanderstraeten}, \citenamefont {Haegeman},\ and\ \citenamefont {Verstraete}}]{vanderstraeten2019tangent}%
  \BibitemOpen
  \bibfield  {author} {\bibinfo {author} {\bibfnamefont {L.}~\bibnamefont {Vanderstraeten}}, \bibinfo {author} {\bibfnamefont {J.}~\bibnamefont {Haegeman}},\ and\ \bibinfo {author} {\bibfnamefont {F.}~\bibnamefont {Verstraete}},\ }\bibfield  {title} {\bibinfo {title} {Tangent-space methods for uniform matrix product states},\ }\href@noop {} {\bibfield  {journal} {\bibinfo  {journal} {SciPost Physics Lecture Notes}\ ,\ \bibinfo {pages} {007}} (\bibinfo {year} {2019})}\BibitemShut {NoStop}%
\bibitem [{\citenamefont {Temme}\ and\ \citenamefont {Verstraete}(2010)}]{temme2010stochastic}%
  \BibitemOpen
  \bibfield  {author} {\bibinfo {author} {\bibfnamefont {K.}~\bibnamefont {Temme}}\ and\ \bibinfo {author} {\bibfnamefont {F.}~\bibnamefont {Verstraete}},\ }\bibfield  {title} {\bibinfo {title} {Stochastic matrix product states},\ }\href@noop {} {\bibfield  {journal} {\bibinfo  {journal} {Physical review letters}\ }\textbf {\bibinfo {volume} {104}},\ \bibinfo {pages} {210502} (\bibinfo {year} {2010})}\BibitemShut {NoStop}%
\bibitem [{\citenamefont {Eckart}\ and\ \citenamefont {Young}(1936)}]{eckart1936approximation}%
  \BibitemOpen
  \bibfield  {author} {\bibinfo {author} {\bibfnamefont {C.}~\bibnamefont {Eckart}}\ and\ \bibinfo {author} {\bibfnamefont {G.}~\bibnamefont {Young}},\ }\bibfield  {title} {\bibinfo {title} {The approximation of one matrix by another of lower rank},\ }\href@noop {} {\bibfield  {journal} {\bibinfo  {journal} {Psychometrika}\ }\textbf {\bibinfo {volume} {1}},\ \bibinfo {pages} {211} (\bibinfo {year} {1936})}\BibitemShut {NoStop}%
\bibitem [{\citenamefont {Ferris}\ and\ \citenamefont {Vidal}(2012)}]{ferris2012perfect}%
  \BibitemOpen
  \bibfield  {author} {\bibinfo {author} {\bibfnamefont {A.~J.}\ \bibnamefont {Ferris}}\ and\ \bibinfo {author} {\bibfnamefont {G.}~\bibnamefont {Vidal}},\ }\bibfield  {title} {\bibinfo {title} {Perfect sampling with unitary tensor networks},\ }\href@noop {} {\bibfield  {journal} {\bibinfo  {journal} {Physical Review B—Condensed Matter and Materials Physics}\ }\textbf {\bibinfo {volume} {85}},\ \bibinfo {pages} {165146} (\bibinfo {year} {2012})}\BibitemShut {NoStop}%
\bibitem [{\citenamefont {Crisanti}\ \emph {et~al.}(2012)\citenamefont {Crisanti}, \citenamefont {Paladin},\ and\ \citenamefont {Vulpiani}}]{crisanti2012products}%
  \BibitemOpen
  \bibfield  {author} {\bibinfo {author} {\bibfnamefont {A.}~\bibnamefont {Crisanti}}, \bibinfo {author} {\bibfnamefont {G.}~\bibnamefont {Paladin}},\ and\ \bibinfo {author} {\bibfnamefont {A.}~\bibnamefont {Vulpiani}},\ }\href@noop {} {\emph {\bibinfo {title} {Products of random matrices: in Statistical Physics}}},\ Vol.\ \bibinfo {volume} {104}\ (\bibinfo  {publisher} {Springer Science \& Business Media},\ \bibinfo {year} {2012})\BibitemShut {NoStop}%
\bibitem [{\citenamefont {Bulchandani}\ \emph {et~al.}(2024)\citenamefont {Bulchandani}, \citenamefont {Sondhi},\ and\ \citenamefont {Chalker}}]{bulchandani2024random}%
  \BibitemOpen
  \bibfield  {author} {\bibinfo {author} {\bibfnamefont {V.~B.}\ \bibnamefont {Bulchandani}}, \bibinfo {author} {\bibfnamefont {S.}~\bibnamefont {Sondhi}},\ and\ \bibinfo {author} {\bibfnamefont {J.}~\bibnamefont {Chalker}},\ }\bibfield  {title} {\bibinfo {title} {Random-matrix models of monitored quantum circuits},\ }\href@noop {} {\bibfield  {journal} {\bibinfo  {journal} {Journal of Statistical Physics}\ }\textbf {\bibinfo {volume} {191}},\ \bibinfo {pages} {55} (\bibinfo {year} {2024})}\BibitemShut {NoStop}%
\bibitem [{\citenamefont {Oseledets}(1968)}]{oseledets1968multiplicative}%
  \BibitemOpen
  \bibfield  {author} {\bibinfo {author} {\bibfnamefont {V.~I.}\ \bibnamefont {Oseledets}},\ }\bibfield  {title} {\bibinfo {title} {A multiplicative ergodic theorem. characteristic ljapunov, exponents of dynamical systems},\ }\href@noop {} {\bibfield  {journal} {\bibinfo  {journal} {Trudy Moskovskogo Matematicheskogo Obshchestva}\ }\textbf {\bibinfo {volume} {19}},\ \bibinfo {pages} {179} (\bibinfo {year} {1968})}\BibitemShut {NoStop}%
\end{thebibliography}%

\clearpage
\newpage
\begin{widetext}
\makeatletter
\begin{center}
\textbf{\large Supplemental Material -- Diverging conditional correlation lengths in the approach to high temperature}
\end{center}
\setcounter{secnumdepth}{2}

\title{}
\author{Jerome Lloyd}
\affiliation{Department of Theoretical Physics, University of Geneva, Geneva, Switzerland}

\author{Dmitry A. Abanin}
\affiliation{Google Quantum AI, Santa Barbara CA, USA}
\affiliation{Department of Physics, Princeton University, Princeton NJ 08544, USA}

\author{Sarang Gopalakrishnan}
\affiliation{Department of Electrical Engineering, Princeton University, Princeton NJ 08544, USA}

\maketitle

\section{Classical distributions as matrix-product states}\label{app:mps}

In this appendix, we describe the formalism for representing classical probability distributions as matrix-product states (MPS).
This analytical framework is used in the rest of the appendices; we defer details of our numerical MPS simulations until Appendix \ref{app:numerics}. While MPS can be used to represent arbitrary probability distributions, they are especially useful for 1d local systems, where the locality of interactions often leads to efficient representations. In the rest of this section, we assume spins $\sigma_i$ arranged in a 1d chain.

We consider the probability distribution over $L$ binary random variables (spins) $\Sigma_i$:
\begin{equation}\label{eq:distribution}
    \pi(\sigma_1,\ldots\sigma_L) = P(\Sigma_1=\sigma_1,\ldots,\Sigma_L=\sigma_L),
\end{equation}
or $\pi({\vec{\sigma}})$ for short. We will often use letters $A,B,\ldots,$ to label a subset of spins e.g.~$A = [a_{1},\ldots,a_{|A|}]$, and $\bar{A}$ to denote the complement. The marginal probability distribution on $A$ is denoted by
\begin{equation}
    \pi_A(\vec{\sigma}_a) = \sum_{\sigma_i:i\in \bar{A}} \pi({\vec{\sigma}}). 
\end{equation}
Likewise, the conditional distribution of $A$ given $B$ is denoted
\begin{equation}
    \pi_{A|B}(\vec{\sigma}_a|\vec{\sigma}_b) = \frac{\pi_{AB}(\vec{\sigma}_a,\vec{\sigma}_b)}{\pi_B(\vec{\sigma}_b)}.
\end{equation}
When there is no ambiguity we will write e.g. $\pi(\vec{\sigma}_a|\vec{\sigma}_b)$ in place of $\pi_{A|B}(\vec{\sigma}_a|\vec{\sigma}_b)$. The distribution in Eq.~(\ref{eq:distribution}) can be written in the form of a matrix-product state as follows: for each position $i$ we define a rank-3 tensor $X^{[i]\sigma_i}_{ab}$, where $\sigma_i$ is the spin (`physical') index with dimension $d=2$, and $a,b$ are left and right `virtual' indices, with dimensions $D_L$ and $D_R$ respectively. We will only consider the translationally invariant case, in which case we write $X^{[i]\sigma_i}_{ab} = X^{\sigma_i}_{ab}$, with dimensions $d\times D\times D$. Then, the distribution $\pi$ is defined as 
\begin{equation}
    \pi(\sigma_1,\ldots,\sigma_L) = \text{Tr}[X^{\sigma_1}\ldots X^{\sigma_L}] \equiv \bra{v_L} X^{\sigma_1}\ldots X^{\sigma_L}\ket{v_R},
\end{equation}
where the internal virtual indices are summed over, and $v_{L,R}$ are $D$-dimensional boundary vectors. The boundary vectors are normalised so that the total distribution satisfies $\sum_{\vec{\sigma}} \pi(\vec{\sigma}) = 1$. The constant $D$ is known as the `bond' dimension, and controls the total amount of correlations contained in the distribution: all `product states' (independent probability distributions) can be written with bond dimension $D=1$.

In the remainder of the paper, we consider the thermodynamic limit $L\to \infty$: in this case there is some ambiguity in defining the boundary vectors and state normalisation. However, all physical observables are calculated with respect to the marginal $\pi_A$ on some reduced region $A$ of spins. In this case, we can introduce the transfer matrix 
\begin{equation}
    T_{ab} = \sum_\sigma X^\sigma_{ab}.
\end{equation}
We assume that the largest eigenvalue of $T$, $\lambda_0$, is unique and w.l.o.g.~we may take $\lambda_0=1$ by renormalising the $X$ tensors. (This is not true for the case of the Ising ground states, as we will comment below.) The corresponding left and right eigenvectors are $\bra{V_L}$ and $\ket{V_R}$. Then, after tracing out the spins in the left and right complement of $A$, we have 
\begin{equation}
    \pi_A(\vec{\sigma}_a) = \bra{V_L} X^{\sigma_{a,1}} \ldots X^{\sigma_{a,{L_a}}} \ket{V_R},
\end{equation}
which is properly normalised assuming $\bra{V_L}V_R\rangle=1$. 

\subsubsection{Example: thermal MPS for 1d Ising model}

As a useful example, consider the 1d Ising model $H(\vec{\sigma}) = -\sum_i \sigma_i\sigma_{i+1}$. The thermal distribution at inverse temperature $\beta$ is $\theta_\beta(\vec{\sigma}) \propto e^{-\beta H(\vec{\sigma})}$. We can write this concisely as an MPS with bond dimension $D=2$:
\begin{equation}\label{eq:thermalmps}
    X^{\sigma=1} = \begin{pmatrix} e^\beta & e^{-\beta} \\ 0 & 0
    \end{pmatrix}, \hspace{1cm} X^{\sigma = -1} = \begin{pmatrix} 0 & 0 \\ e^{-\beta} & e^{\beta} 
    \end{pmatrix}.
\end{equation}
The transfer matrix 
\begin{equation}
    T = \begin{pmatrix} e^\beta & e^{-\beta} \\ e^{-\beta} & e^{\beta}
    \end{pmatrix}
\end{equation}
is the standard one for the 1d Ising model.

\subsubsection{}


\section{Conditional mutual information for ground state depolarization}\label{app:groundstate}

In this appendix, we calculate the conditional mutual information in the case of depolarizing noise acting on the Ising ground states. From (\ref{eq:thermalmps}), the initial distribution takes the form  
\begin{equation}
X(t=0) = 
\left(
\begin{array}{cc}
\left(\begin{smallmatrix} 1 \\ 0 \end{smallmatrix}\right) & 0 \\
0 & \left(\begin{smallmatrix} 0 \\ 1 \end{smallmatrix}\right)
\end{array}
\right),\label{eq:gstensors}
\end{equation}
where we use the inner vector space to label the physical spin space. The depolarizing noise channel acts independently on each spin,  and corresponds to contracting the $A$ tensors with the noise matrix in Eq.~(\ref{eq:depolchannel}):
\begin{equation}\label{eq:depolmps}
    X^{\sigma_i}(t) = \sum_{\sigma'_i}\mathcal{E}_{p(t)}^{(i)}(\sigma_i|\sigma_i')X^{\sigma'_i}(0), \hspace{1cm} X(t) = \left(
\begin{array}{cc}
\left(\begin{smallmatrix} q \\ p \end{smallmatrix}\right) & 0 \\
0 & \left(\begin{smallmatrix} p \\ q \end{smallmatrix}\right)
\end{array}
\right),
\end{equation}
with $p \equiv p(t)$ and $q = 1-p$. Note that, due to the symmetry of the initial state (preserved by the noise channel), the transfer matrix $T(t)$ is equal to the identity at all times: therefore the leading eigenvalue is non-unique, and the leading eigenvectors are not uniquely specified. This extra degree of freedom corresponds to the relative weights of the two symmetry sectors: for the fully symmetric state  we take $\ket{V_L} = \ket{V_R}=  \begin{pmatrix}
    1 & 1
\end{pmatrix}/\sqrt{2}$. The marginal distribution on subregion $B$ is given by the `symmetrised' Bernouilli distribution:
\begin{equation}\label{eq:bernoulli}
    \pi(\vec{\sigma}_b;t) = \frac{1}{2} q ^{L_+}p ^{L_-} + \frac{1}{2} p ^{L_+}q ^{L_-},
\end{equation}
where $L_+$ denotes the number of spins taking a value of $+1$, and $L_+ + L_- = |B|$. We will assume for simplicity $|B|$ even and set $L_+ = |B|/2+l$, $L_- = |B|/2-l$. 

In order to calculate the conditional mutual information, we employ the sum formula Eq.~(\ref{eq:CMIsum}). We assume that $A = [a]$ and $C=[c]$ each represent a single spin. We first calculate the conditional distribution $\pi({\sigma}_a,{\sigma}_c|\vec{\sigma}_b;t)$. This can be found by contracting the MPS to be
\begin{equation}\label{eq:conditional}
    \pi({\sigma}_a,{\sigma}_c|\vec{\sigma}_b;t) = \begin{pmatrix}
        \frac{q^2\lambda^{-l}+p^2\lambda^{l}}{\lambda^l+\lambda^{-l}} & pq \\
        pq & \frac{p^2\lambda^{-l}+q^2\lambda^{l}}{\lambda^l+\lambda^{-l}}
    \end{pmatrix} = pq \begin{pmatrix}
        \frac{\cosh ((l+1)\log \lambda)}{\cosh (l\log \lambda)} & 1 \\
        1 & \frac{\cosh ((l-1)\log \lambda)}{\cosh (l\log \lambda)}
    \end{pmatrix},
\end{equation}
where $\lambda = \frac{p}{q}$. Now, since both the marginal $\pi(\vec{\sigma}_b;t) \equiv \pi(l;t)$ and the conditional distribution $\pi({\sigma}_a,{\sigma}_c|\vec{\sigma}_b;t) \equiv \pi({\sigma}_a,{\sigma}_c|l;t) $ depend only on $\vec{\sigma}_b$ through the number $l$, the CMI simplifies to the sum 
\begin{equation}
    I_t(A:C|B) = \sum_{l = -|B|/2}^{|B|/2} \pi(l;t) I_{t}(A:C|l). 
\end{equation}
where $I_t(A:C|l)$ is calculated on $\pi({\sigma}_a,{\sigma}_c|l;t)$. 
Importantly, to evaluate the above equation we only need to sum over $|B|$ values of the mutual information (as opposed to $2^{|B|}$), and each $I_t(A:C|l)$ can be computed explicitly from Eq.~(\ref{eq:conditional}) using the standard formula $I(A:C) = S_A+S_B-S_{AB}$. The CMI can therefore be exactly computed even for large values of $|B|$, giving the results in Fig.~\ref{fig:glauber}a. 

n the asymptotic late-time limit, the CMI can be further evaluated:
We write $z = -\log \lambda $, with $z \to 0^+$ as $t \to \infty$. Note that, in order to extract the Markov length, we are interested in the order of limits $\lim_{t\to\infty} \lim_{|B|\to\infty}$, and therefore must be careful how we handle the limit $lz $ --- setting $z = 0$ in Eq.~(\ref{eq:conditional}) gives vanishing mutual information for all configurations $\vec{\sigma}_b$. Instead, we decompose the conditional distribution according to
\begin{equation}\label{eq:conditional2}
    \pi(\sigma_a,\sigma_c|\vec{\sigma}_b;t) =  \frac{e^{-z}}{(1+e^{-z})^2} \begin{pmatrix}
        \cosh z +\sinh z\tanh zl  & 1 \\
        1 &  \cosh z -\sinh z\tanh zl
    \end{pmatrix}.
\end{equation}
Then, for $z\ll1$, the mutual information between $A$ and $C$, conditioned on measuring $L_+ = \frac{L}{2}+l$ $B$-spins up, can be expanded using $I(A:C) = S_A+S_C-S_{AC}$, which after a little algebra yields 
\begin{equation}
    I_{\pi(t)}(A:C|l) = \frac{z^4}{32}\text{sech}^4 (zl).
\end{equation}
Next, using the Central Limit Theorem, we rewrite the Bernoulli distribution Eq.~(\ref{eq:bernoulli}) as 
\begin{equation}\label{eq:symnormal}
    \pi(l;t) = \frac{1}{2}\mathcal{N}(l;\mu_+,v)+\frac{1}{2}\mathcal{N}(l;\mu_-,v),
\end{equation}
where 
\begin{equation}
    \mathcal{N}(l,\mu,v) = \frac{1}{\sqrt{2\pi v^2}} e^{-\frac{(l-\mu)^2}{2v^2}},
\end{equation}
is Gaussian and $\mu_\pm = \pm(\frac{1}{2}-p)|B| = \pm\frac{|B|}{2}\tanh \frac{z}{2}$, $ v^2 = pq|B| = \frac{e^{-z}}{(1+e^{-z})^2}|B| $. The distribution in Eq.~(\ref{eq:symnormal}) is rather special: in the late-time limit when $z \ll 1$, the two Gaussian peaks have standard deviation $v \sim \frac{\sqrt{|B|}}{2}$ and are separated by a distance $\sim \frac{|B|z}{2}$. To accurately resolve the individual peaks, we require $z^2 |B| \gg 1$, which is the CLT prediction for the number of samples needed to accurately resolve the initial magnetisation. 

Combining the above formulas, and in the limits $z\ll 1$, $z^2|B| \gg 1$, the CMI is given by the integral formula
\begin{gather}
  I_{\pi(t)}(A:C|B) \approx \frac{z^3}{32\sqrt{2\pi v^2}} \int_{-\infty}^\infty dx\ g(x), \hspace{.5cm} g(x)=\frac{1}{2}\bigg(e ^{-\frac{(x-z\mu_+)^2}{2v^2z^2}}+e ^{-\frac{(x-z\mu_-)^2}{2v^2z^2}}\bigg)\text{sech}^4 x.
\end{gather}
The maximum of the function $g(x)$ occurs at $x = 0$, $g(0) = e^{-\mu_+^2/2v^2} \approx e^{-z^2|B|/8}$. We can then evaluate the integral via the saddle point method about $x=0$, with the large parameter $z^2|B|$, leading to 
\begin{equation}
    \int_{-\infty}^\infty dx\ g(x) \approx e^{-z^2|B|/8} \sqrt{\frac{2\pi}{3+\frac{4}{z^2|B|}}}.
\end{equation}
Combining, we have the CMI at leading order in $z^2 |B| $ as given by Eq.~(\ref{eq:collapse}), i.e.
\begin{equation}
    I_{\pi(t)}(A:C|B) \approx \frac{z^4}{16\sqrt{3z^2|B|}}e^{-z^2|B|/8}.\label{eq:asymcmi}
\end{equation}
The late-time Markov length is therefore
\begin{equation}
    \xi_M(t) = \frac{8}{z^2} = \frac{2}{m^2},
\end{equation}
where $m = \tanh\frac{z}{2} \approx \frac{z}{2}$ is the late-time magnetisation. Thus we see that for the Ising ground states evolving under depolarizing noise, the late-time Markov length is set by the CLT scaling $\xi_M \propto \frac{1}{m^2}$. This scaling is confirmed in Fig.~\ref{fig:glauber}(a). 


\section{Exact formula for spin-spin correlation function under Glauber dynamics}\label{app:spinspin}

In the main text, we considered the Glauber dynamics defined by the local discrete-time update rules, Eqs.~(\ref{eq:thermalupdate}-\ref{eq:glauberweights2}). The continuous-time version of this dynamics was introduced in Glauber's original paper \cite{glauber1963time}, where he used it to compute the evolution of low-order spin correlators undergoing thermalising dynamics. The dynamics for the two-point correlator 
\begin{equation}
    C_r(t) = \sum_{\vec{\sigma}} \sigma_0 \sigma_r \pi(\vec{\sigma};t),
\end{equation}
is of particular interest, and additionally serves as a good benchmark for our MPS simulations (see Section \ref{app:numerics}). The exact calculation for $C_r(t)$ requires a few minor modifications to Glauber's solution on account of the discrete-time setup, which we detail below.

We first derive the evolution of the on-site magnetisation, $m_j(t) = \sum_{\vec{\sigma}} \sigma_j \pi(\vec{\sigma};t)$, which serves as a warm-up to the two-point case. To this end, we rewrite the transition matrix elements, Eq.~(\ref{eq:glauberweights}), as follows:

\begin{equation}
    W^{\sigma,\sigma'}_{\tau,\tau'}(\beta) = (1-\alpha)\delta_{\sigma,\sigma'}+\alpha \frac{\exp(\beta \sigma(\tau+\tau'))}{\sum_{\sigma''} \exp(\beta \sigma''(\tau+\tau'))} = (1-\alpha)\delta_{\sigma,\sigma'}+\frac{\alpha}{2}(1+\frac{\gamma}{2}\sigma(\tau'+\tau)),
\end{equation}
where $\gamma = \tanh 2\beta$. Next, assuming first $j$ even, we find that under the action of $\Phi^{\text{even}}_\beta$ the magnetisation evolves to 
\begin{gather}
    m_j(t+1/2)= \sum_{\vec{\sigma}\vec{\sigma}'} \sigma_j  \Phi^{\text{even}}_\beta(\vec{\sigma}|\vec{\sigma}')\pi(\vec{\sigma}';t) =\sum_{\vec{\sigma}\vec{\sigma}'\vec{\tau}} \sigma_j  \prod_{i\ \text{even}}  \Delta_{\tau_{i-1},\tau_{i}}^{\sigma_{i-1},\sigma'_{i-1}}W^{\sigma_i,\sigma'_i}_{\tau_i,\tau_{i+1}}(\beta) \pi(\vec{\sigma}';t) \nonumber \\
    = \sum_{\sigma_j,\vec{\sigma}'} \sigma_j W^{\sigma_j,\sigma'_j}_{\sigma'_{j-1}\sigma'_{j+1}} \pi(\vec{\sigma}';t) \nonumber \\ 
    = \sum_{\vec{\sigma}'}\big[(1-\alpha)\sigma_j'+\frac{\alpha\gamma}{2}(\sigma_{j-1}'+\sigma'_{j+1})]\pi(\vec{\sigma}';t) \nonumber \\
    = (1-\alpha)m_{j}(t)+\frac{\alpha\gamma}{2}(m_{j-1}(t)+m_{j+1}(t)).\label{eq:magupdate}
\end{gather}
The application of the $\Phi^{\text{odd}}_\beta$ channel does not change the magnetisation further, and hence $m_j(t+1) = m_j(t+1/2)$. For $j$ odd, we have instead 
\begin{equation}
    m_j(t+1) = (1-\alpha)m_{j}(t)+\frac{\alpha\gamma}{2}(m_{j-1}(t+1/2)+m_{j+1}(t+1/2)).
\end{equation}
Writing $\alpha = dt$ and $dm_j(t)=m_j(t+1)-m_j(t)$, we recover in the limit $\alpha \to 0$ Glauber's solution:
\begin{equation}
    \frac{dm_{j}(t)}{dt} = -m_{j}(t) +\frac{\gamma}{2}(m_{j-1}(t)+m_{j+1}(t)).
\end{equation}
For uniform initial magnetisation, $m_j(0) = m$, giving the expected exponentially decaying solution $m_j(t) = e^{-(1-\gamma)t}m$.

Next, consider the two-spin correlation function $C_r(t)$, where we assume translational invariance. Assuming $r$ even for simplicity, under $\Phi^\text{even}_\beta$ the correlator evolves as
\begin{gather}
    C_{r}(t+1/2) = \sum_{\vec{\sigma}\vec{\sigma}'} \sigma_0\sigma_r  W^{\sigma_0,\sigma'_0}_{\sigma'_{-1}\sigma'_{1}}W^{\sigma_r,\sigma'_r}_{\sigma'_{r-1}\sigma'_{r+1}}\pi(\vec{\sigma}';t)  \nonumber \\ 
   = \sum_{\vec{\sigma}'} \big[(1-\alpha)^2\sigma'_0\sigma'_r+\frac{(1-\alpha)\alpha\gamma}{2}(\sigma'_r(\sigma'_{-1}+\sigma'_1)+\sigma'_0(\sigma'_{r-1}+\sigma'_{r+1}) +\frac{\alpha^2\gamma^2}{4}(\sigma'_{-1}+\sigma'_1)(\sigma'_{r-1}+\sigma'_{r+1}) ]\pi(\vec{\sigma}') \nonumber \\
   = 
    (1-2\alpha+\alpha^2(1+\gamma^2/2)) C_r(t) +(1-\alpha)\gamma \alpha(C_{m-1}(t)+C_{m+1}(t))+\frac{\alpha^2\gamma^2}{4}(C_{m-2}(t)+C_{m+2}(t)),\label{eq:spinspin}
\end{gather}
for $r\neq 0$. For $r=0$, we have the identity $C_0(t) = 1$. Again applying $\Phi^\text{odd}_\beta$ does not change the function further and $C_r(t+1) = C_r(t+1/2)$. Notice that this equation, under the replacement $C_r(t) \to m_r(t)$, is nothing but the `square' of the equation for the magnetisation, Eq.~(\ref{eq:magupdate}): the only difference compared to the magnetisation's evolution is the fixed boundary condition, $C_0(t) = 1$, which leads to correlations decaying in the distance $r$. For $\alpha \to 0$ we recover Glauber's solution again:
\begin{equation}\label{eq:corrupdate}
    \frac{dC_r(t)}{dt} = -2C_r(t)+\gamma(C_{r-1}(t)+C_{r+1}(t)).
\end{equation}
For finite $\alpha$ the quadratic correction appears. It can be readily checked that the ansatz $C_r = e^{-|r|/\xi_\beta}$, where $\xi_\beta = -1/\log\tanh\beta$ in the thermal correlation length, is a steady state solution to Eq.~(\ref{eq:corrupdate}): for the Glauber dynamics starting from initial finite temperature, $C_r(0) = e^{-|r|/\xi_{\beta_i}}$ also sets the initial condition. Eq.~(\ref{eq:spinspin}) can be efficiently iterated to obtain the time-dependent spin-spin correlators (exact solutions expressed in terms of the Bessel functions are obtained in the small $\alpha$ limit in \cite{glauber1963time}).

\section{Thermal mutual information for the 1d Ising model}

In the main text, we argued that, for initially thermal states quenched to higher temperatures $\beta _f< \beta_i$, the Markov length of the late-time conditional mutual information is set by the decay of the disconnected mutual information (between $A$ and $C$) in the original thermal state. The aim of this section is to derive the thermal mutual information, and show that it decays with a decay length $\xi_{\beta_i}/2$. 

We use the MPS representation of the thermal state, Eq.~(\ref{eq:thermalmps}), and consider $A$ and $C$ to each consist of a single spin, separated by a contiguous region of $|B|$ spins. In order to calculate the mutual information $I(A:C)$, we diagonalise the transfer matrix to derive the marginal $\theta_{\beta_i}(\sigma_a,\sigma_c)$:
\begin{gather}
    \theta_{\beta_i}(\sigma_a,\sigma_c) = \sum_{k=1,2} \lambda_k^{|B|} \bra{V_L}X^{\sigma_a}\ket{V_k}\bra{V_k}X^{\sigma_c}\ket{V_R} \nonumber \\
    = \frac{1}{4}\begin{pmatrix}
        1+\lambda_2^{|B|+1} & 1-\lambda_2^{|B|+1} \\
        1-\lambda_2^{|B|+1} & 1+\lambda_2^{|B|+1},
    \end{pmatrix}
\end{gather}
where $\ket{V_k}$ are the eigenvectors of $T$, $\ket{V_1} \equiv \ket{V_R}$, $\bra{V_1} \equiv \bra{V_L}$, and $\lambda_2 = \tanh \beta_i$. To calculate the MI, we use $I(A:C) = S_A+S_C-S_{AC}$. The entropies for the marginals on $A$ and $C$ are clearly both equal to one bit. For the joint entropy, we have 
\begin{gather}
    S_{AC} = -\frac{1+\lambda_2^{|B|+1}}{2}\log \frac{1+\lambda_2^{|B|+1}}{4} - \frac{1-\lambda_2^{|B|+1}}{2}\log \frac{1+\lambda_2^{|B|+1}}{4} \nonumber \\
    = -\frac{1+\lambda_2^{|B|+1}}{2}\log \frac{1+\lambda_2^{|B|+1}}{2} - \frac{1-\lambda_2^{|B|+1}}{2}\log \frac{1-\lambda_2^{|B|+1}}{2} + 1 \nonumber \\ 
    = 1+H_2(\zeta(|B|)), 
\end{gather}
where we defined the binary entropy function $H_2(\zeta) = -\zeta \log \zeta-(1-\zeta)\log(1-\zeta)$, and $\zeta(|B|) = \frac{1}{2}(1+(\tanh\beta_i)^{|B|})$. Putting it all together, we find the thermal mutual information 
\begin{equation}
    I_{\theta_{\beta_i}}(A:C) = 1- H_2(\zeta(|B|)).
\end{equation}
Interestingly, this is equal to the channel capacity for the binary symmetric channel, with a probability of error $\zeta(|B|)$: this has a natural interpretation as $B$ acts as a lossy transmission channel between $A$ and $C$, with information loss caused by spontaneous magnetisation fluctuations.    

For $\beta_i$ finite and large $|B|$, we can expand the binary entropy in the small parameter $\epsilon = (\tanh\beta_i)^{|B|} \ll 1$, 
\begin{equation}
    H_2(\zeta \to 1/2) = 1-\frac{\epsilon^2}{2\ln2} + \mathcal{O}(\epsilon^4).
\end{equation}
This gives the limit of the MI for large separation of $A$ and $C$:
\begin{equation}
    I_{\theta_{\beta_i}}(A:C) \approx \frac{(\tanh\beta_i)^{2|B|}}{2\ln 2} = \frac{e^{-2|B|/\xi_{\beta_i}}}{2\ln 2}. 
\end{equation}
By conflating the decay lengths of the thermal MI and the late-time CMI, we find the Markov length in the late-time limit of Glauber dynamics is given by 
\begin{equation}
    \xi_M = \frac{\xi_{\beta_i}}{2}.
\end{equation}

\section{Details of MPS simulations}\label{app:numerics}

Our MPS simulations use the basic formalism described in Section \ref{app:mps} of the SM. Since we work with translationally invariant MPS in the thermodynamic limit, we only need to store a single evolving tensor $X(t)$, which has dimensions $d\times D\times D$. In this case the simulation cost is independent of the system size. The initial state of the dynamics is given by Eq.~(\ref{eq:thermalmps}), which has $D=2$, for initial temperature $\beta_i$. To implement the time evolution, we first write the Glauber dynamics in the form of a matrix product operator (MPO), described below, and then employ a version of the ``Time Evolving Block Decimation'' (TEBD) method to propagate the system in time \cite{vidal2004efficient, verstraete2004matrix}. In its standard formulation, TEBD is used to simulate unitary time-evolution of quantum wavefunctions; here, we simulate stochastic evolution of \emph{classical} distribution functions. This requires some care, namely re-finding the so-called canonical form of the MPS after each time-evolution step, as we now explain.

To write the noise channel as an MPO, we first introduce the tensors corresponding to Eqs.~(\ref{eq:glauberweights}, \ref{eq:glauberweights2}), using standard graphical notation:

\begin{gather}
W_{\tau,\tau'}^{\sigma,\sigma'}(\beta) = 
    \begin{tikzpicture}
        [baseline = (Y.base),every node/.style={scale=0.75},scale=.5]
\draw (0.5,2.5) -- (1,2.5); 
\draw (1,3) rectangle (2,2);
\draw (1.5,2.5) node (Y) {$W$};
\draw (2,2.5) -- (2.5,2.5); 
\draw (1.5,3) -- (1.5,3.5);
\draw (1.5,2) -- (1.5,1.5);
\draw (1.5,1.5) node[anchor=north] (s) {$\sigma;$};
\draw (1.5,3.5) node[anchor=south] (sp) {$\sigma$};
\draw (0.5,2.5) node[anchor= east] (t) {$\tau$};
\draw (2.5,2.5) node[anchor= west] (tp) {$\tau'$};
\end{tikzpicture} 
= (1-\alpha)\delta_{\sigma,\sigma'}+\alpha \frac{\exp(\beta \sigma(\tau+\tau'))}{\sum_{\sigma''} \exp(\beta \sigma''(\tau+\tau'))}, 
\\
\Delta_{\tau,\tau'}^{\sigma,\sigma'} = 
\begin{tikzpicture}
    [baseline = (Y.base),every node/.style={scale=0.75},scale=.55]
\draw (0.5,2.5) -- (1,2.5); 
\fill[gray] (1,3) rectangle (2,2);
\draw (2,2.5) -- (2.5,2.5); 
\draw (1.5,3) -- (1.5,3.5);
\draw (1.5,2) -- (1.5,1.5);
\draw (1.5,1.5) node[anchor=north] (s) {$\sigma'$};
\draw (1.5,3.5) node[anchor=south] (sp) {$\sigma$};
\draw (0.5,2.5) node[anchor= east] (t) {$\tau$};
\draw (2.5,2.5) node[anchor= west] (tp) {$\tau'$};
\end{tikzpicture} 
= \delta_{\sigma,\sigma'}\delta_{\tau,\tau'}\delta_{\sigma,\tau'}.
\end{gather}
The Glauber noise channel is then written as they two layer MPO (first acting on even spins, then odd):

\begin{equation}
\Phi_\beta = 
      \begin{tikzpicture}[baseline={(current bounding box.center)},scale=0.5]

[baseline = (Y.base),every node/.style={scale=0.75},scale=.5]
\draw (1.5,2) -- (1.5,2.5); \draw (3.5,2) -- (3.5,2.5); \draw (5.5,2) -- (5.5,2.5);
\draw (7.5,2) -- (7.5,2.5); \draw (9.5,2) -- (9.5,2.5);

\draw (0.5,3) -- (1,3); 
\fill[gray] (1,3.5) rectangle (2,2.5);
\draw (1.5,3.5) -- (1.5,4);

\draw (2,3) -- (3,3); 
\draw (3,3.5) rectangle (4,2.5);
\draw (3.5,3) node (Y) {$W$};
\draw (3.5,3.5) -- (3.5,4);

\draw (4,3) -- (5,3); 
\fill[gray] (5,3.5) rectangle (6,2.5);
\draw (5.5,3.5) -- (5.5,4);

\draw (6,3) -- (7,3); 
\draw (7,3.5) rectangle (8,2.5);
\draw (7.5,3) node {$W$};
\draw (7.5,3.5) -- (7.5,4);

\draw (8,3) -- (9,3); 
\fill[gray] (9,3.5) rectangle (10,2.5);
\draw (10,3) -- (10.5,3); 
\draw (9.5,3.5) -- (9.5,4);

\draw (0.5,4.5) -- (1,4.5); 
\draw (1,5) rectangle (2,4);
\draw (1.5,4.5) node {$W$};
\draw (1.5,5) -- (1.5,5.5);

\draw (2,4.5) -- (3,4.5); 
\fill[gray] (3,5) rectangle (4,4);

\draw (3.5,5) -- (3.5,5.5);

\draw (4,4.5) -- (5,4.5); 
\draw (5,5) rectangle (6,4);
\draw (5.5,4.5) node {$W$};
\draw (5.5,5) -- (5.5,5.5);

\draw (6,4.5) -- (7,4.5); 
\fill[gray] (7,5) rectangle (8,4);
\draw (7.5,5) -- (7.5,5.5);

\draw (8,4.5) -- (9,4.5); 
\draw (9,5) rectangle (10,4);
\draw (10,4.5) -- (10.5,4.5); 
\draw (9.5,5) -- (9.5,5.5);
\draw (9.5,4.5) node {$W$};

\end{tikzpicture}.\label{eq:glaubermpo}
\end{equation}
The MPO has bond dimension $D_{\text{Gl}}=4$. At each time step we have $\pi(t+1) = \Phi_\beta(\pi(t))$, where the MPO is contracted with the MPS over the physical legs (lower legs in diagram (\ref{eq:glaubermpo})) -- note that in practice we only need to perform the contraction on a single copy of $X$, due to the translational symmetry. To account for the even-odd effect, we perform the contraction over a two-site unit cell.

The procedure described so far is exact, but leads to an exponentially growing MPS bond dimension $D \sim D_{\text{Gl}}^t$. Therefore, after each time evolution step, we truncate the MPS bond dimension back into the manifold of fixed bond dimension $D\leq D_{\text{max}}$. To ensure that the effect of truncation errors is minimal, it is essential that the MPS is written in the so-called `canonical form' before truncating \cite{orus2008infinite}. This amounts to a gauge transformation acting at the virtual level
\begin{equation}
    X_{\text{Y}} = YXY^{-1},
\end{equation}
for invertible $Y$.
Note that, while we deal with a classical MPS in our work, the canonical form is the \emph{same} as for \emph{quantum} systems: following \cite{vanderstraeten2019tangent}, we define the left orthonormal form $X_L = L X L^{-1}$ via the identity
\begin{equation}
    \sum_{\sigma,a} (X_L)_{ab}^\sigma (\bar{X}_L)_{ab'}^{\sigma} = \delta_{b,b'},
\end{equation}
where $\bar{X}$ denotes the complex conjugate. Similarly, we define the right orthonormal form $X_R = R^{-1} X R$ via 
\begin{equation}
    \sum_{\sigma,b} (X_R)_{ab}^\sigma (\bar{X}_R)_{a'b}^{\sigma} = \delta_{a,a'}.
\end{equation}
We then define the \emph{orthogonality center} $X_C$ according to 
\begin{equation}
    X_C = X_LC = CX_R, \hspace{1cm} C = LR.
\end{equation}
Now, a generic infinite MPS can be gauged as 
\begin{equation} \pi = \hspace{.2cm}
\begin{tikzpicture}[baseline = (X.base),every node/.style={scale=0.75},scale=.5]
\draw (0.5,1.5) -- (1,1.5); 
\draw[rounded corners] (1,2) rectangle (2,1);
\draw (1.5,1.5) node (X) {$X$};
\draw (2,1.5) -- (3,1.5); 
\draw[rounded corners] (3,2) rectangle (4,1);
\draw (3.5,1.5) node {$X$};
\draw (4,1.5) -- (5,1.5);
\draw[rounded corners] (5,2) rectangle (6,1);
\draw (5.5,1.5) node {$X$};
\draw (6,1.5) -- (7,1.5); 
\draw[rounded corners] (7,2) rectangle (8,1);
\draw (7.5,1.5) node {$X$};
\draw (8,1.5) -- (8.5,1.5); 
\draw (1.5,2) -- (1.5,2.5); \draw (3.5,2) -- (3.5,2.5); \draw (5.5,2) -- (5.5,2.5);
\draw (7.5,2) -- (7.5,2.5);
\end{tikzpicture} \hspace{.2cm}=\hspace{.2cm}
\begin{tikzpicture}[baseline = (X.base),every node/.style={scale=0.75},scale=.5]
\draw (0.5,1.5) -- (1,1.5); 
\draw[rounded corners] (1,2) rectangle (2,1);
\draw (1.5,1.5) node (X) {$X_L$};
\draw (2,1.5) -- (3,1.5); 
\draw[rounded corners] (3,2) rectangle (4,1);
\draw (3.5,1.5) node {$X_L$};
\draw (4,1.5) -- (5,1.5);
\draw[rounded corners] (5,2) rectangle (6,1);
\draw (5.5,1.5) node {$C$};
\draw (6,1.5) -- (7,1.5); 
\draw[rounded corners] (7,2) rectangle (8,1);
\draw (7.5,1.5) node {$X_R$};
\draw (8,1.5) -- (9,1.5); 
\draw[rounded corners] (9,2) rectangle (10,1);
\draw (9.5,1.5) node {$X_R$};
\draw (10,1.5) -- (10.5,1.5);
\draw (1.5,2) -- (1.5,2.5); \draw (3.5,2) -- (3.5,2.5); 
\draw (7.5,2) -- (7.5,2.5); \draw (9.5,2) -- (9.5,2.5);
\end{tikzpicture}.
\end{equation}
Finding the matrices $L$, $R$ defining the canonical forms is the computationally expensive part of the time-evolution step. We use the algorithm based on iterated QR decompositions detailed in Section 2.3 of Ref.~\cite{vanderstraeten2019tangent}, which has a cost scaling as $O(dD^3)$. Alternatively, the method of Ref.~\cite{orus2008infinite} can be used with the same cost scaling, but care must be taken to avoid inverting small values. 

There exists a remaining freedom in the definitions of the left and right canonical forms, up to a unitary transform i.e.~ $X_L' = U^\dagger X_LU$. This is taken advantage of in order to write the matrix $C$ in diagonal form by performing a singular-value decomposition: $C = UC'V^\dagger$. Henceforth we drop the primes since there is no confusion. Due to the left and right canonical forms, the bipartition across the orthogonality center now defines a Schmidt decomposition of the state,
\begin{equation}
    \pi_{[D']} = \sum_{\lambda = 1}^{D'} C_\lambda \Psi_L^{\lambda}(X_L) \Psi^\lambda_R(X_R),
\end{equation}
where the $\Psi_L^{\lambda}(X_L)$ ($\Psi_R^{\lambda}(X_R)$) form an orthonormal basis (w.r.t.~$\lambda$) on the left and right halves of the chain, the $C_\lambda$ are assumed to be ordered $C_{i+1} \leq C_i$, and we use $\pi_{[D]}$ to denote the sum restricted to the first $D$ singular values. We then truncate the Schmidt values $C_\lambda$ by keeping only the largest $D \leq D'$ values (for numerical stability we also discard small values below some cutoff, $C \leq \delta$, where in our numerics we used $\delta = 10^{-9}$): $\pi^{[D']} \to \pi^{[D]}$. Note that the $\Psi^{\lambda}$ above do \emph{not} represent valid states (marginals) in the classical case, since they generally contain negative valued elements. For an alternative construction where the elements are restricted to real non-negative values, see \cite{temme2010stochastic}: the disadvantage of this approach is that the singular-value decomposition is replaced by a matrix factorisation routine known as non-negative matrix factorisation, for which there is currently no known algorithm for finding exact solutions \cite{johnson2010dynamical}. 

In summary, one time-evolution step consists of the following steps:
\begin{enumerate}
    \item Contract the (two-site) MPS tensor $X(t)$ with the Glauber MPO tensors, $W$ and $\Delta$, according to Eq.~(\ref{eq:glaubermpo}). This gives the tensor $\tilde{X}(t)$, having bond dimension $D'$.
    \item Find the left and right canonical forms, $\tilde{X}_L$ and $\tilde{X}_R$, and the diagonal Schmidt matrix $C$, using e.g.~the algorithm of Ref.~\cite{vanderstraeten2019tangent}.  
    \item Truncate the tensors to a maximum bond dimension $D$, by discarding the smallest $\text{max}(0, D'-D)$ Schmidt values. The truncated left canonical matrix can be used as the new MPS tensor, $\tilde{X}(t)_{[D]} = X(t+1)$. 
\end{enumerate}

We note that, while the truncation based on the Schmidt spectrum is optimal for finite quantum systems, in the sense that it maximises the overlap between the original and truncated states for a given bipartition and bond dimension $D$ (this follows from the Eckart-Young theorem \cite{eckart1936approximation}), it is not expected to necessarily be optimal for classical distributions \cite{johnson2010dynamical}. Nevertheless, for small singular value truncation, we expect the truncated distribution to closely approximate the true distribution. In order to check the accuracy of the MPS simulations, we compare the value of the spin-spin correlation function, computed from the MPS, to the exact dynamics derived in Section \ref{app:spinspin} of the SM. We compute the correlator of the MPS from the formula
\begin{equation}
    C_r(t) = \sum_{\sigma \sigma'} \sigma \sigma'  \bra{V_L}X^\sigma T^{(r-1)}X^\sigma\ket{V_R}.
\end{equation}
We fix the bond dimension $D=18$ and plot the results in Fig.~\ref{fig:benchmark} for three different combinations of the initial and final temperatures. The correlation function calculated from the MPS matches the exact result extremely well, only starting to deviate for late times and large distances when the magnitude of the correlation function becomes small $\lesssim 10^{-7}$. The temperatures in Fig.~\ref{fig:benchmark}a,c correspond to the dataset used in the main text, Fig.~\ref{fig:glauber}, and we conclude that our MPS is accurate in the regimes studied. 

\begin{figure}
    \centering
    \includegraphics[width=.9\linewidth]{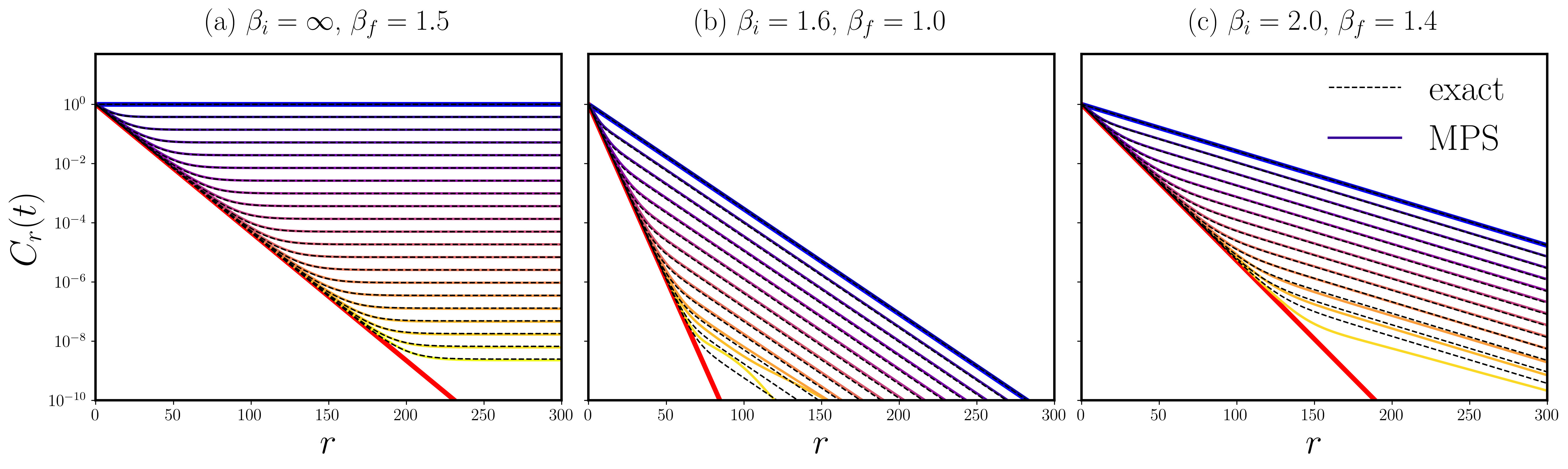}
    \caption{\emph{MPS benchmark:} comparison of spin-spin correlator $C_r(t)$, computed from the time-evolved MPS (coloured lines), and the exact results Eq.~(\ref{eq:spinspin}) (dotted black lines). Colour transitions from blue to yellow indicate increasing time, with thick blue (red) lines the thermal state results at zero (infinite) times. We show data for three different initial and final temperatures with (a) and (c) corresponding to the data in Fig.~\ref{fig:glauber} in the main text. We use a bond dimension $D=18$ and a value $\alpha = 1$ in (a) and $\alpha=0.5$ in (b,c). The MPS result matches the exact results extremely well, up until late times and large distances where the value of the correlation function is below $\sim10^{-8}$. }
    \label{fig:benchmark}
\end{figure}

When the initial state is taken as the ground state, the MPS becomes non-injective (the leading eigenvalue of the transfer matrix is degenerate), and the state exhibits long-range order. The Ising $\mathbb{Z}_2$ symmetry implies that there exists a symmetry at the virtual level, $X^\sigma = VX^{-\sigma}V^{-1}$, which leaves the global state invariant under a spin flip on all sites --- for the ground state tensors, Eq.~(\ref{eq:gstensors}), we have $V=\sigma_x$, where $\sigma_x$ is the Pauli-$x$ matrix. The variational method for finding the canonical forms typically does not preserve this structure, even though the Glauber dynamics preserves the symmetry. To get around this, for our ground state simulations we perform the time evolution starting from the polarized state $X^\sigma = \delta_{\sigma,0}$. When we want access to the symmetric state (e.g.~for computing observables), we form the block state 
\begin{equation}
    \tilde{X}^\sigma = \begin{pmatrix}
        X^\sigma & 0 \\0 & X^{-\sigma} 
    \end{pmatrix} \hspace{.5cm}\longrightarrow \hspace{.5cm}\tilde{X}^\sigma = V\tilde{X}^{-\sigma}V^{-1}, \hspace{.5cm} V = \begin{pmatrix}
    0 & \mathds{1}_{D\times D} \\
    \mathds{1}_{D\times D} & 0 
    \end{pmatrix}.
\end{equation}
On account of $V$, the global state obtained from $\tilde{X}^\sigma$ is symmetric. 

\section{Method for computing conditional mutual information from MPS }

In this subsection, we explain our algorithm to sample the conditional mutual information from MPS. The zero temperature and finite temperature cases are treated slightly differently: we first explain the finite temperature case, which is the generic one, and then discuss the modification needed when starting from the ground state.

We assume translational invariance and an MPS $\pi$ described by a local tensor $X^\sigma_{ab}$. Extension to the non-translationally invariant case is straightforward. The CMI is defined by Eq.~(\ref{eq:CMIsum}). Our method for calculating the CMI combines the exact sampling algorithm for MPS \cite{ferris2012perfect} to generate configurations of $\vec{\sigma}_b$ of the $B$ region, with the fact that the MI for the conditioned state $\pi(\sigma_a,\sigma_c|\vec{\sigma}_b;t)$ can be efficiently evaluated. 

First, for given $X^\sigma_{ab}$, we compute the transfer matrix $T$, normalised with leading eigenvalue $\lambda_0=1$, and the leading left and right eigenvectors $\bra{V_{L}}$, $\ket{V_R}$, normalised according to $\bra{V_L}V_R\rangle=1$. For later use we also precompute the contractions of the left and right environments with the tensors for the $A$ and $B$ sites: $\bra{A^{\sigma_a}} = \bra{V_L}X^{\sigma_a}$, $\ket{C^{\sigma_c}} = X^{\sigma_c}\ket{V_R}$. This can easily be generalised to the case where $A$ and $C$ contain more than a single spin, in which case we have e.g.~$\bra{A^{\sigma_{a,1},\sigma_{a,2}\ldots\sigma_{a,k}}} = \bra{V_L}X^{\sigma_{a,1}}X^{\sigma_{a,2}}\ldots X^{\sigma_{a,k}}$. 

Perfect sampling for MPS stems from the chain rule for probabilities:
\begin{equation}
    \pi(\sigma_1,\sigma_2,\ldots,\sigma_L) = \pi(\sigma_1)\pi(\sigma_2|\sigma_1)\ldots\pi(\sigma_L|\sigma_{L-1},\ldots,\sigma_1).
\end{equation}
We first form the single-site marginal (we drop the $b$ labels for simplicity)
\begin{equation}
    \pi(\sigma_1) = \bra{V_L}X^{\sigma_1}\ket{V_R}.
\end{equation}
We then randomly choose a value for $\sigma_1$ with the corresponding weights (i.e.~draw a random number $0\leq r\leq1$ and compare to the cumulative probability distribution of $\pi(\sigma_1)$). We denote the specific value of $\sigma_1$ by $\hat \sigma _1 =1,\ldots,d$, and set $F(\hat\sigma_1) = X^{\hat \sigma _1}/\pi(\hat \sigma _1)$. Note that $F$ is a $D\times D$ matrix. We then compute the conditional distribution:
\begin{equation}
    \pi(\sigma_2|\hat \sigma_1) = \bra{V_L}F(\hat \sigma_1)X^{\sigma_2}\ket{V_R}
\end{equation}
and likewise randomly sample $\hat\sigma_2$. We then set $F(\hat\sigma_1,\hat\sigma_2) = X^{\hat \sigma _1}X^{\hat \sigma _2}/\pi(\hat\sigma_1,\hat\sigma_2) = F(\hat\sigma_1)X^{\hat \sigma _2}/\pi(\hat\sigma_2|\hat \sigma _1)$ and $\pi(\hat\sigma_1,\hat\sigma_2) = \pi(\hat\sigma_1)\pi(\hat\sigma_2|\hat\sigma_1)$.
We then proceed sampling the value of $\hat \sigma_k$ from 
\begin{equation}
    \pi(\sigma_k|\hat\sigma_{k-1}\ldots\hat\sigma_1) = \bra{V_L}F(\hat\sigma_1,\ldots,\hat\sigma_{k-1})X^{\sigma_k}\ket{V_R},
\end{equation}
with 
\begin{equation}
    F(\hat\sigma_1,\ldots,\hat\sigma_{k}) = \frac{X^{\hat\sigma_1}X^{\hat\sigma_2}\ldots X^{\hat\sigma_k}}{\pi(\hat\sigma_1,\hat\sigma_2,\ldots \hat\sigma_k)} = \frac{F(\hat\sigma_1,\ldots,\hat\sigma_{k-1})X^{\hat\sigma_k}}{\pi(\hat \sigma_k|\hat\sigma_{k-1}\ldots\hat\sigma_1)}.
\end{equation}
At the end of the sampling process we have generated one sample of the pair $[\pi(\hat\sigma_1,\ldots,\hat\sigma_L), F(\hat{\sigma}_1,\ldots\hat\sigma_L)]$, with corresponding probability $\pi(\hat{\sigma}_1,\ldots\hat\sigma_L)$. We may then form the conditional marginal distribution on $A$ and $C$ via
\begin{equation}
    \pi(\sigma_a,\sigma_c|\hat{\sigma}_1,\ldots\hat\sigma_L) = \bra{A^{\sigma_a}}F(\hat{\sigma}_1,\ldots\hat\sigma_L)|C^{\sigma_c}\rangle.
\end{equation}
On account of the matrix multiplications the cost of generating one sample of $\pi(\sigma_a,\sigma_c|\hat{\sigma}_1,\ldots\hat\sigma_L)$ is $dLD^3$, and requires drawing $L$ random numbers. Note that $\pi(\sigma_a,\sigma_c|\hat{\sigma}_1,\ldots\hat\sigma_L)$ is a normalised distribution over $d\times d$ values and we can compute the MI via the standard formula, i.e.
\begin{equation}
    I(\sigma_a,\sigma_c|B=\hat{\sigma}_1,\ldots\hat\sigma_L) = \sum_{\sigma_a,\sigma_c}\pi(\sigma_a,\sigma_c|\hat{\sigma}_1,\ldots\hat\sigma_L)\log \bigg[\frac{\pi(\sigma_a,\sigma_c|\hat{\sigma}_1,\ldots\hat\sigma_L)}{\pi(\sigma_a|\hat{\sigma}_1,\ldots\hat\sigma_L)\pi(\sigma_c|\hat{\sigma}_1,\ldots\hat\sigma_L)}\bigg].\label{eq:CMIrelent}
\end{equation}
After generate $Q$ samples, we form an unbiased estimate for the CMI as 
\begin{equation}
    \bar I(A:C|B) = \frac{1}{Q}\sum_{q=1}^Q I(\sigma_a,\sigma_c|B=\hat{\sigma}^q_1,\ldots\hat\sigma^q_L).
\end{equation}

The above algorithm is summarised in the pseudocode below:
\begin{algorithm}[H]

\caption{Sample conditional mutual information from MPS}
\begin{spacing}{1.3}
\begin{algorithmic}[1]
\vspace{.3cm}
\Require $Q$ samples, $L\ge1$ 
\For{$q=1,\ldots,Q$}
\State $\Pi \gets 1$, $F \gets \mathds{1}_{D\times D}$
\For{$k=1,\ldots L$}

\State $\pi(\sigma_k|\hat\sigma_{k-1}\ldots\hat\sigma_1) \gets  \bra{V_L} F(\hat\sigma_1,\ldots,\hat\sigma_{k-1})X^{\sigma_k}\ket{V_R}$
\State draw $\hat\sigma_k$ with probability $\pi(\sigma_k|\hat\sigma_{k-1}\ldots\hat\sigma_1)$
\State $\Pi \gets \Pi \times \pi(\hat\sigma_k|\hat\sigma_{k-1}\ldots\hat\sigma_1)$, $F\gets F\ X^{\hat\sigma_k}/\pi(\hat\sigma_k|\hat\sigma_{k-1}\ldots\hat\sigma_1)$
\EndFor
\State $\pi(\sigma_a,\sigma_c|\hat\sigma_1,\ldots,\hat\sigma_L) \gets \bra{A^{\sigma_a}}F(\hat{\sigma}_1,\ldots\hat\sigma_L)|C^{\sigma_c}\rangle$
\State compute $I(\sigma_a,\sigma_c|B=\hat{\sigma}_1,\ldots\hat\sigma_L)$ from $\pi(\sigma_a,\sigma_c|\hat\sigma_1,\ldots,\hat\sigma_L)$ using Eq.~(\ref{eq:CMIrelent}).
\EndFor
\State $\bar I(A:C|B) \gets \frac{1}{Q}\sum_{q=1}^Q I(\sigma_a,\sigma_c|B=\hat{\sigma}^q_1,\ldots\hat\sigma^q_L)$
\end{algorithmic}
\end{spacing}
\end{algorithm}

In the case of ground state heating, due to the non-injectivity of the MPS we perform the time evolution starting from the fully polarized state, as described in the previous section. In order to explicitly enforce the symmetrisation in the CMI calculation, we modify the algorithm as follows: as well as the matrix $F$, we track the matrix
\begin{equation}
    \bar{F}(\hat\sigma_1,\ldots,\hat\sigma_{k}) = \frac{\bar F(\hat\sigma_1,\ldots,\hat\sigma_{k-1})\bar{X}^{\hat\sigma_k}}{\pi(\hat \sigma_k|\hat\sigma_{k-1}\ldots\hat\sigma_1)},
\end{equation}
where $\bar{X}^\sigma = X^{-\sigma}$. Then, in calculating the MI $I(\sigma_a,\sigma_c|B=\hat{\sigma}_1,\ldots\hat\sigma_L)$, we use the symmetrised marginal:
\begin{equation}
    \tilde{\pi}(\sigma_a,\sigma_c|\hat{\sigma}_1,\ldots\hat\sigma_L) = \frac{1}{2}\bra{A^{\sigma_a}}F(\hat{\sigma}_1,\ldots\hat\sigma_L)|C^{\sigma_c}\rangle + \frac{1}{2}\bra{A^{-\sigma_a}}\bar F(\hat{\sigma}_1,\ldots\hat\sigma_L)|C^{-\sigma_c}\rangle. 
\end{equation}


\section{Additional numerical data }

In this section we present data supplementary to the main text. First we investigate the time-dependent scaling of the Markov length, starting from the symmetric Ising ground state and evolving under the Glauber dynamics with a final temperature $\xi_M$. In the main text, we argued for the case of infinite temperature noise that the Markov length behavior should be governed by the scaling $\xi_M(t) \sim 1/m^2(t)$, in terms of the time-dependent magnetisation. This reasoning is still expected to hold for the finite temperature noise case, since the initial ground state polarization can be reconstructed by majority vote decoding for all finite times. However, the central limit theorem applies to independently identically distributed spins. Since spins are decorrelated on distances the scale of the thermal length $\xi_{\beta_f}$, a naive coarse-graining argument predicts an enhancement of the Markov length for ground states evolving under finite temperature noise:
\begin{equation}\label{eq:groundthermalscaling}
    \xi_M(t) \sim \frac{\xi_{\beta_f}}{m^2(t)}.
\end{equation}
In Fig.~\ref{fig:markovlengths}(left), we plot the numerically extracted Markov lengths (points) for four final temperatures, and compare with the prediction Eq.~(\ref{eq:groundthermalscaling}). We observe good agreement for late times, and conclude that the Markov length grows exponentially as in the depolarization case, but with a different pre-factor. 

Next, we confirm the late-time scaling of the Markov length starting from a thermal state, which is expected to go as 
\begin{equation}
    \xi_M \sim \frac{1}{2}\xi_{\beta_i},
\end{equation}
from the arguments presented in the main text. Note that the final Markov length is independent of the final state temperature. We confirm this scaling for a range of initial and final temperatures in Fig.~\ref{fig:markovlengths}(right). 

\begin{figure}
    \centering
    \includegraphics[width=0.8\linewidth]{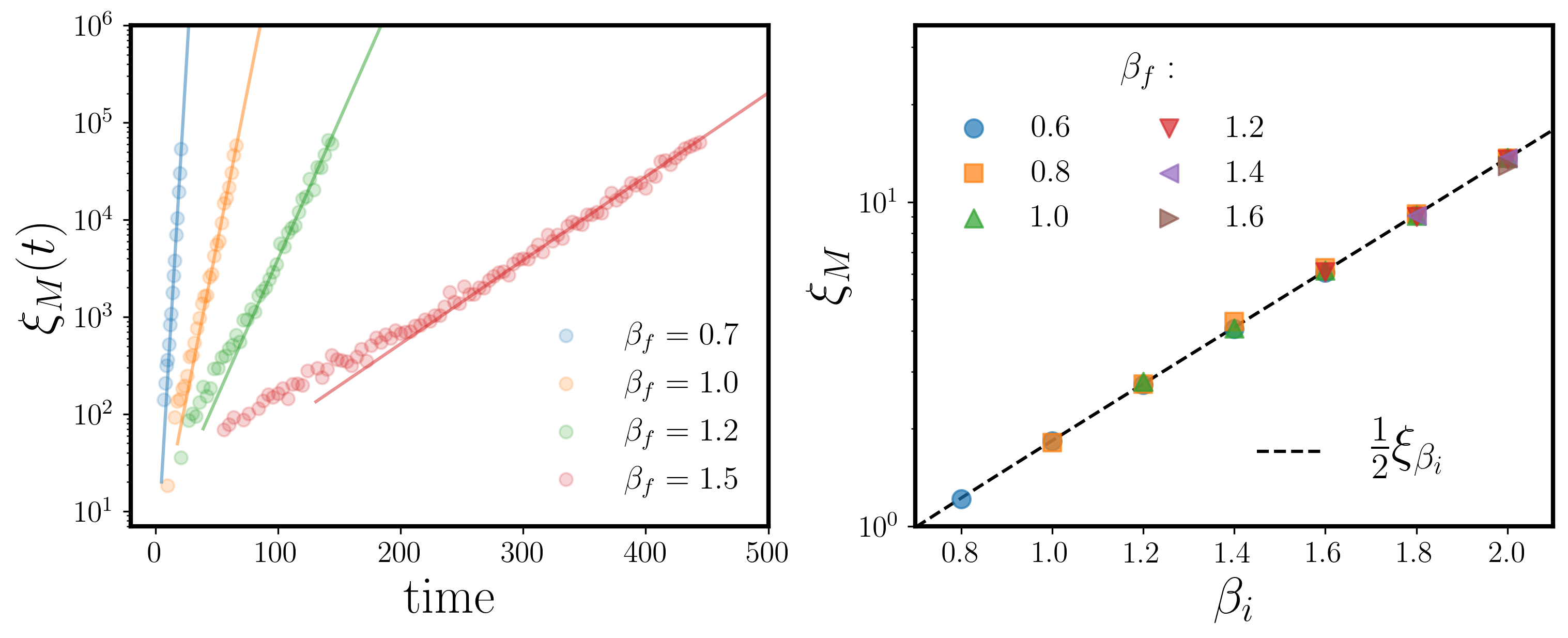}
    \caption{\emph{Left:} growth of time-dependent Markov length (points), starting from the ground state and under Glauber noise with final inverse temperature $\beta_f$. The solid lines are the predictions $\xi_M(t) = \xi_{\beta_f}/m^2(t)$ based on the CLT, where $m(t)$ is the time-dependent magnetisation (starting from the polarized ground state). \emph{Right}: Late-time Markov lengths for finite temperature quenches, $\beta_i \to \beta_f$. The late-time length is in agreement with the thermal predition $\frac{1}{2}\xi_{\beta_i}$ (black dashed line), and is independent of initial temperature.}
    \label{fig:markovlengths}
\end{figure}

We finally comment on the relation between the Markov length and the Lyapunov spectrum \cite{crisanti2012products, bulchandani2024random}. To define the Lyapunov spectrum, we consider the random matrix 
\begin{equation}
        F(\hat\sigma_1,\ldots,\hat\sigma_{k}) = \frac{X^{\hat\sigma_1}X^{\hat\sigma_2}\ldots X^{\hat\sigma_k}}{\pi(\hat\sigma_1,\hat\sigma_2,\ldots \hat\sigma_k)},
\end{equation}
used in the calculation of the CMI. We denote by $\lambda_k$ the singular values of $F$. The Lyapunov spectrum is defined as 
\begin{equation}
    \eta_k = \lim_{|B|\to\infty} \frac{\log \lambda_k}{|B|}.
\end{equation}
The importance of the Lyapunov spectrum can be stated in the form of Oseledets' ergodic theorem \cite{oseledets1968multiplicative}, which under mild conditions states that the $\eta_k$ are in fact non-random with probability one. Thus, for large enough $|B|$, the singular values for typical realisations of $F$ are fixed by the Lyapunov spectrum. For large $|B|$, we write the matrix $F$ in terms of the largest two singular values, where we assume $\lambda_0$ is unique and $|\lambda_1|<|\lambda_0|$:
\begin{equation}
    F \sim \lambda_0[\ket{r_0(\vec{\sigma}_b)}\bra{l_0(\vec{\sigma}_b)}+\frac{\lambda_1}{\lambda_0}\ket{r_1(\vec{\sigma}_b)}\bra{l_1(\vec{\sigma}_b)}],
\end{equation}
where $\ket{r_k(\vec{\sigma}_b)}$, $\bra{l_k(\vec{\sigma}_b)}$ are the right/left singular vectors. In the limit where $\lambda_1/\lambda_0$ goes to zero, the marginal $\tilde{\pi}(\sigma_a,\sigma_c|\hat{\sigma}_1,\ldots\hat\sigma_L)$ factorizes into the product distribution:
\begin{equation}
    \pi(\sigma_a,\sigma_c|\hat{\sigma}_1,\ldots\hat\sigma_L) = \lambda_0\langle A^{\sigma_a}\ket{r_0(\vec{\sigma}_b)}\bra{l_0(\vec{\sigma}_b)}C^{\sigma_c}\rangle \equiv \pi(\sigma_a|\vec{\sigma}_b)\pi(\sigma_c|\vec{\sigma}_b).
\end{equation}
For small $\lambda_1/\lambda_0$, we may write 
\begin{equation}
    \pi(\sigma_a,\sigma_c|\hat{\sigma}_1,\ldots\hat\sigma_L) =\pi(\sigma_a|\vec{\sigma}_b)\pi(\sigma_c|\vec{\sigma}_b) + e^{-w|B|} g(\sigma_a,\sigma_c),
\end{equation}
where $w=\eta_0-\eta_1$ and $\sum_{\sigma_a,\sigma_c}g(\sigma_a,\sigma_c) = 0$. To leading order in the small parameter $e^{-w|B|}$, a Taylor expansion shows that the mutual information scales as $I(\sigma_a,\sigma_c|B=\hat{\sigma}_1,\ldots\hat\sigma_L) \propto e^{-2w|B|}$, with the first order term vanishing. Since the Lyapnuov parameter $w$ is independent of the random configuration $\vec{\sigma}_b$, we conclude that the Markov length scales as
\begin{equation}\label{eq:lyap}
    \xi_M(t) = \frac{1}{2w(t)}.
\end{equation}
In Fig.~(\ref{fig:lyapunov}), we compare the time-dependent Markov length (points) for initial temperature $\beta_i = 2$ and several final temperatures, to the value predicted from the Lyapunov spectrum, Eq.~(\ref{eq:lyap}) (lines). We find strong agreement between the two results in all cases.

\begin{figure}
    \centering
    \includegraphics[width=0.5\linewidth]{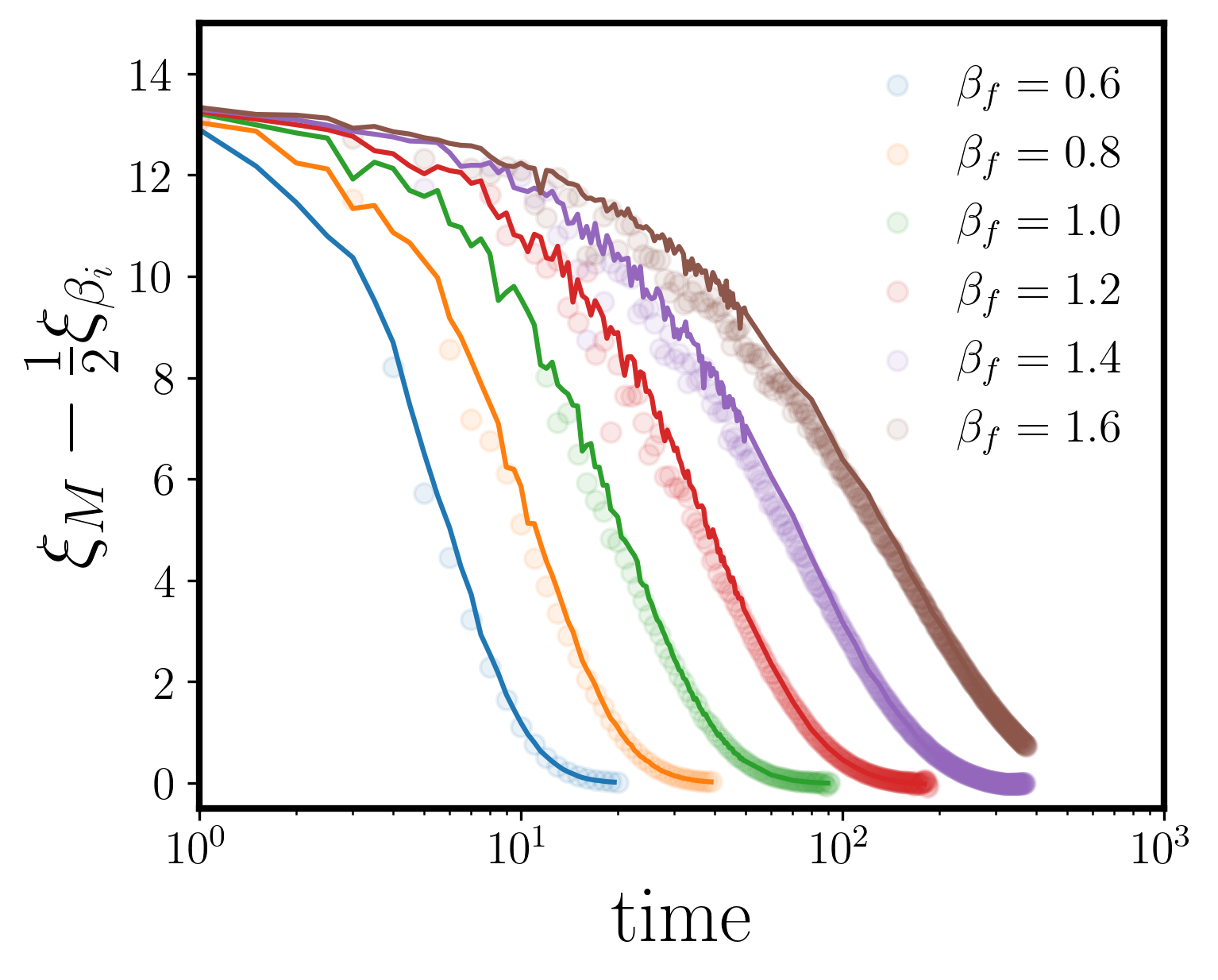}
    \caption{Time-dependent Markov length (points) compared to Lyapunov exponent prediction, Eq.~(\ref{eq:lyap}) (lines). We show results for $\beta_i=2$ and several final temperatures, and plot the Markov length relative to the late time value $\frac{1}{2}\xi_{\beta_i}$. }
    \label{fig:lyapunov}
\end{figure}

\end{widetext}

\end{document}